\documentstyle[11pt]{article}
\newcounter{subequation}[equation]

\makeatletter

\def\thesubequation{\theequation\@alph\c@subequation}
\def\@subeqnnum{{\rm (\thesubequation)}}
\def\slabel#1{\@bsphack\if@filesw {\let\thepage\relax
   \xdef\@gtempa{\write\@auxout{\string
      \newlabel{#1}{{\thesubequation}{\thepage}}}}}\@gtempa
   \if@nobreak \ifvmode\nobreak\fi\fi\fi\@esphack}
\def\subeqnarray{\stepcounter{equation}
\let\@currentlabel=\theequation\global\c@subequation\@ne
\global\@eqnswtrue
\global\@eqcnt\z@\tabskip\@centering\let\\=\@subeqncr
$$\halign to \displaywidth\bgroup\@eqnsel\hskip\@centering
  $\displaystyle\tabskip\z@{##}$&\global\@eqcnt\@ne
  \hskip 2\arraycolsep \hfil${##}$\hfil
  &\global\@eqcnt\tw@ \hskip 2\arraycolsep
  $\displaystyle\tabskip\z@{##}$\hfil
   \tabskip\@centering&\llap{##}\tabskip\z@\cr}
\def\endsubeqnarray{\@@subeqncr\egroup
                     $$\global\@ignoretrue}
\def\@subeqncr{{\ifnum0=`}\fi\@ifstar{\global\@eqpen\@M
    \@ysubeqncr}{\global\@eqpen\interdisplaylinepenalty \@ysubeqncr}}
\def\@ysubeqncr{\@ifnextchar [{\@xsubeqncr}{\@xsubeqncr[\z@]}}
\def\@xsubeqncr[#1]{\ifnum0=`{\fi}\@@subeqncr
   \noalign{\penalty\@eqpen\vskip\jot\vskip #1\relax}}
\def\@@subeqncr{\let\@tempa\relax
    \ifcase\@eqcnt \def\@tempa{& & &}\or \def\@tempa{& &}
      \else \def\@tempa{&}\fi
     \@tempa \if@eqnsw\@subeqnnum\refstepcounter{subequation}\fi
     \global\@eqnswtrue\global\@eqcnt\z@\cr}
\let\@ssubeqncr=\@subeqncr
\@namedef{subeqnarray*}{\def\@subeqncr{\nonumber\@ssubeqncr}\subeqnarray}
\@namedef{endsubeqnarray*}{\global\advance\c@equation\m@ne%
                           \nonumber\endsubeqnarray}

\makeatletter
\@addtoreset{equation}{section}
\makeatother
\renewcommand{\theequation}{\thesection.\arabic{equation}}

\def\dalemb#1#2{{\vbox{\hrule height .#2pt
        \hbox{\vrule width.#2pt height#1pt \kern#1pt
                \vrule width.#2pt}
        \hrule height.#2pt}}}

\def\half{{\textstyle{1\over2}}}
\let\a=\alpha \let\b=\beta   \let\e=\epsilon
  \let\q=\theta  
  \let\n=\nu

\let\la=\label  
  
\def\nn{\nonumber} \def\bd{\begin{document}} \def\ed{\end{document}}
\def\ds{\documentstyle} \let\fr=\frac \let\bl=\bigl \let\br=\bigr
\let\Br=\Bigr \let\Bl=\Bigl
\let\bm=\bibitem
\let\na=\nabla
\let\pa=\partial \let\ov=\overline
\def\ie{{\it i.e.\ }}
\newcommand{\be}{\begin{equation}}
\newcommand{\ee}{\end{equation}}
\def\ba{\begin{array}}
\def\ea{\end{array}}
\def\ft#1#2{{\textstyle{{\scriptstyle #1}\over {\scriptstyle #2}}}}
\def\fft#1#2{{#1 \over #2}}
\def\del{\partial}
\def\sst#1{{\scriptscriptstyle #1}}
\def\oneone{\rlap 1\mkern4mu{\rm l}}
\def\e7{E_{7(+7)}}
\def\td{\tilde}
\def\wtd{\widetilde}
\def\im{{\rm i}}
\def\bog{Bogomol'nyi\ }
\def\q{{\tilde q}}
\def\hast{{\hat\ast}}
\def\0{{\sst{(0)}}}
\def\1{{\sst{(1)}}}
\def\2{{\sst{(2)}}}
\def\3{{\sst{(3)}}}
\def\4{{\sst{(4)}}}
\def\5{{\sst{(5)}}}
\def\6{{\sst{(6)}}}
\def\7{{\sst{(7)}}}
\def\8{{\sst{(8)}}}
\def\n{{\sst{(n)}}}
\def\oo{{\"o}}
\def\hA{\hat{\cal A}}
\def\ns{{\sst {\rm NS}}}
\def\rr{{\sst {\rm RR}}}
\def\tH{{\widetilde H}}
\def\tB{{\widetilde B}}
\def\cA{{\cal A}}
\def\cF{{\cal F}}
\def\tF{{\wtd F}}
\def\Z{\rlap{\sf Z}\mkern3mu{\sf Z}}
\def\ep{{\epsilon}}
\def\IIA{{\rm IIA}}
\def\IIB{{\rm IIB}}
\def\ads{{\rm AdS}}
\def\R{\rlap{\rm I}\mkern3mu{\rm R}}

\def\Ei{{\hbox{Ei}}}
\def\Ci{{\hbox{Ci}}}
\def\Si{{\hbox{Si}}}

\newcommand{\ho}[1]{$\, ^{#1}$}
\newcommand{\hoch}[1]{$\, ^{#1}$}
\newcommand{\bea}{\begin{eqnarray}}
\newcommand{\eea}{\end{eqnarray}}
\newcommand{\ra}{\rightarrow}
\newcommand{\lra}{\longrightarrow}
\newcommand{\Lra}{\Leftrightarrow}
\newcommand{\ap}{\alpha^\prime}
\newcommand{\bp}{\tilde \beta^\prime}
\newcommand{\tr}{{\rm tr} }
\newcommand{\Tr}{{\rm Tr} }
\newcommand{\NP}{Nucl. Phys. }
\newcommand{\tamphys}{\it Center for Theoretical Physics,
Texas A\&M University, College Station, TX 77843}
\newcommand{\upenn}{\it Dept. of Phys. and Astro.,
University of Pennsylvania,
Philadelphia, PA 19104}

\newcommand{\auth}{M. Cveti\v{c}\hoch{\dagger1},
M.J. Duff\hoch{\ddagger2}, P. Hoxha\hoch{\ddagger},
James T. Liu\hoch{\star3},
H. L\"u\hoch{\dagger1},\\
J.X. Lu\hoch{\ddagger2}, R. Martinez-Acosta\hoch{\ddagger},
C.N. Pope\hoch{\ddagger4}, H. Sati\hoch{\ddagger}
and T.A. Tran\hoch{\ddagger}}

\begin{document}
\begin{flushright}
\hfill{
UPR/0840-T \\
CTP TAMU-11/99 \\
RU99-4-B\\
March 1999}\\
\hfill{\bf hep-th/9903214}\\
\end{flushright}


\begin{center}

{\large {\bf Embedding AdS Black Holes in Ten and Eleven Dimensions}}

\vspace{20pt}

\auth

\vspace{10pt}
{\hoch{\dagger}\upenn}

\vspace{10pt}
{\hoch{\ddagger}\tamphys}

\vspace{10pt}
{\hoch{\star}{\it Dept. of Phys., The Rockefeller University,
New York, NY 10021} }

\vspace{30pt}

\underline{ABSTRACT}
\end{center}

   We construct the non-linear Kaluza-Klein ans\"atze describing the
embeddings of the $U(1)^3$, $U(1)^4$ and $U(1)^2$ truncations of
$D=5$, $D=4$ and $D=7$ gauged supergravities into the type IIB string
and M-theory.  These enable one to oxidise any associated lower
dimensional solutions to $D=10$ or $D=11$.  In particular, we use these
general ans\"atze to embed the charged AdS$_{5}$, AdS$_{4}$
and AdS$_{7}$ black hole solutions in ten and eleven dimensions.
The charges for the black holes with toroidal horizons may
be interpreted as the angular momenta of D3-branes, M2-branes and
M5-branes spinning in the transverse dimensions, in their near-horizon
decoupling limits. The horizons of the black holes coincide with the
worldvolumes of the branes. The Kaluza-Klein ans\"atze also allow the
black holes with spherical or hyperbolic horizons to be reinterpreted
in $D=10$ or $D=11$.

{\vfill\leftline{}\vfill
\vskip 10pt \footnoterule
{\footnotesize \hoch{1}
Research supported in part by DOE grant DOE-FG02-95ER40893
\vskip  -12pt} \vskip   14pt
{\footnotesize \hoch{2}
Research supported in part by NSF Grant PHY-9722090
\vskip -12pt}  \vskip  14pt
{\footnotesize \hoch{3}
Research supported in part by DOE grant DOE-91ER40651-TASKB
\vskip -12pt} \vskip 14pt
{\footnotesize \hoch{4}
Research supported in part by DOE grant DOE-FG03-95ER40917
\vskip -12pt}  \vskip  14pt
}

\pagebreak
\setcounter{page}{1}

\tableofcontents
\addtocontents{toc}{\protect\setcounter{tocdepth}{2}}
\newpage

\section{Introduction\label{sec:intro}}

Anti-de Sitter black hole solutions of gauged extended supergravities 
\cite{romans} are currently attracting a good deal of attention
\cite{Behrndt1,birm,Caldarelliklemm1,Klemm,Behrndt2,%
Duffliu,Chamblin,Cveticgubser1,%
Cveticgubser2,Sabra,Caldarelliklemm2} due, in large
part, to the correspondence between anti-de Sitter space and conformal
field theories on its boundary
\cite{Maldacena,Gubserklebanovpolyakov,Witten,Witten2}.  These gauged
extended supergravities can arise as the massless modes of various
Kaluza-Klein compactifications of both $D=11$ and $D=10$
supergravities. The three examples studied in the paper will be gauged
$D=4$, $N=8$ $SO(8)$ supergravity \cite{deWit1,deWit2} arising from
$D=11$ supergravity on $S^{7}$ \cite{duffpope3,DNP} whose black hole
solutions are discussed in \cite{Duffliu}; gauged $D=5$, $N=8$
$SO(6)$ supergravity \cite{PPV,GRW} arising from Type IIB supergravity
on $S^{5}$ \cite{Schwarz,GM,KRV} whose black hole solutions are
discussed in \cite{Behrndt1,Behrndt2}; and gauged $D=7$, $N=4$ $SO(5)$
supergravity \cite{PPV,TV} arising from $D=11$ supergravity on $S^{4}$
\cite{PTV} whose black hole solutions are given in section \ref{seven}
and in \cite{Cveticgubser1,lm}.\footnote{BPS black holes arising in
the $SU(2)\times SU(2)$ version of gauged $N=4$ supergravity in $D=4$,
which is the massless sector of the $S^3\times S^3$ compactification
of $N=1$ supergravity in $D=10$, were discussed in
\cite{Klemm}.  These solutions are not asymptotically AdS.}
In the absence of the black holes,  these three AdS
compactifications are singled out as arising from the near-horizon
geometry of the extremal non-rotating M2, D3 and M5 branes
\cite{GT,DGT,GHT,Duffads}. One of our goals will be to embed these
known lower-dimensional black hole solutions into ten or eleven
dimensions, thus allowing a higher dimensional interpretation in terms
of rotating M2, D3 and M5-branes.  

Since these gauged supergravity theories may be obtained by
consistently truncating the massive modes of the full Kaluza-Klein
theories, it follows that all solutions of the lower-dimensional
theories will also be solutions of the higher-dimensional ones
\cite{Duffpope,Pope}. In principle, therefore, once we know the
Kaluza-Klein ansatz for the massless sector, it ought to be
straightforward to read off the higher dimensional solutions.  It
practice, however, this is a formidable task. The correct massless
ansatz for the $S^{7}$ compactification took many years to finalize
\cite{deWitnicolaiwarner,deWitnicolai}, and is still highly implicit,
while for the $S^{5}$ and $S^{4}$ compactifications, the complete
massless ans\"atze are still unknown. For our present purposes, it
suffices to consider truncations of the gauged supergravities to
include only gauge fields in the Cartan subalgebras of the full gauge
groups, namely $U(1)^{4}$, $U(1)^{3}$ and $U(1)^{2}$ for the $S^{7}$,
$S^{5}$ and $S^{4}$ compactifications, respectively. These truncated
theories will admit respectively the 4-charge AdS$_{4}$, 3-charge
AdS$_{5}$ and 2-charge AdS$_{7}$ black hole solutions.

The simplest of the three is perhaps the $D=5$, $N=8$ maximal gauged
supergravity, for which there is a consistent $N=2$ (\ie minimal)
truncation to supergravity coupled to two abelian vector multiplets.
This has the bosonic field content of a graviton, three $U(1)$ gauge
fields and two scalars.  In this paper we obtain the complete
non-linear Kaluza-Klein ansatz for the compactification of $D=10$
Type IIB supergravity on $S^5$, truncated to the $U(1)^3$ Cartan
subgroup of $SO(6)$.

In four dimensions there is a consistent truncation of gauged $N=8$
maximal supergravity to gauged $N=2$ supergravity coupled to three
vector multiplets. The bosonic sector consists of a graviton, four
vectors and three complex scalars, whose real and imaginary parts
correspond to three ``axions'' and three ``dilatons.''
\footnote{Interestingly enough, the ungauged version of this theory
obtained by switching off the gauge coupling and performing some
dualisations, appears in the $T^{2}$ compactification of $D=6$, $N=1$
string theory. The four vectors are the two Kaluza-Klein and two
winding gauge fields, while the three complex scalars $S$, $T$ and $U$
correspond to the axion-dilaton, the Kahler form and complex structure
of the torus. This $STU$ system plays a crucial role in
four-dimensional string/string/string triality
\cite{Duffliurahmfeld1}.  The black hole solutions of this theory
\cite{cvyod,Duffliurahmfeld1}, and
their embedding in ungauged $N=8$ supergravity \cite{lpsol,khor} 
arising from the
$T^{7}$ compactification of $M$-theory as intersections
\cite{cveticts,tsey} are also well known.}
The inclusion of the axions is necessary for providing a consistent
truncation; the full bosonic Lagrangian in this case is obtained in
appendix B.  This truncation corresponds to the $U(1)^4$ Cartan
subgroup of the non-abelian $SO(8)$, for which there exist AdS black
hole solutions with four electric charges \cite{Duffliu}.  While one
would ideally wish to obtain a complete Kaluza-Klein ansatz for the
$N=2$ truncation, in practice the complexity arising from the
inclusion of the axions is considerable.  Thus in the present paper we
omit the axions in the Kaluza-Klein reduction.  This is of course
sufficient for the embedding of the electric black hole solutions in
$D=11$ as they do not involve the axions.

Finally, in seven dimensions, maximal $N=4$ gauged $SO(5)$
supergravity admits a consistent truncation to $N=2$ supergravity,
comprising the metric, a 2-form potential, three vectors and a
dilaton, coupled to a vector multiplet comprising a vector and three
scalars.  We obtain the Kaluza-Klein ansatz for an $S^4$ reduction of
$D=11$ supergravity, including two $U(1)^2$ gauge fields and two
dilatonic scalars.  This is sufficient for the consideration of the
embedding of the $D=7$ black holes in $D=11$.

  Having obtained the explicit Kaluza-Klein reduction ans\"atze, this
allows an investigation of the embedding of the various AdS black
holes of $D=4$, $D=5$ and $D=7$ in the respective higher-dimensional
supergravities.  An important point here is that one must know the
exact Kaluza-Klein reduction ansatz for the reduction of the
supergravity theory itself, and not just for a specific solution, in
order to show that the metric, gauge fields and scalar fields of the
lower-dimensional solution are indeed precisely embeddable in the
higher-dimensional theory.  It is worth remarking, in this regard,
that it is the scalar fields that present most of the subtleties and
complexities in the implementation of the reduction procedure.

Having embedded these black holes in ten or eleven dimensions, an
interesting question then arises as to their higher-dimensional
interpretation.  It was noted some time ago \cite{BDPS2}, in the
context of a ``test'' membrane moving in a fixed AdS$_{4} \times
S^{7}$, that a 4-dimensional BPS state (whose AdS energy is equal to
its electric charge) admits the eleven-dimensional interpretation of
an M2-brane \cite{BST,dust} that is rotating in the extra
dimensions. Moreover, the electric charge is equal to the spin.

Recently there has been an upsurge of activities on the study of
rotating $p$-branes
\cite{Cveticyoum3,Cveticyoum2,Gubser,Csaki,Kraus,Cai,%
Chamblin,Cveticgubser1,Cveticgubser2,klt,rosf,Csaki2,sfetsos}.  In
particular, in \cite{Chamblin} AdS Reissner-Nordstr\"om black holes
({\it i.e.}~the charged black holes without scalars) of AdS
supergavity in $D=4$ and $D=5$ were studied, and shown to be related
to the rotating solutions of M-/string theory.  In
\cite{Cveticgubser1} the near-extreme spinning D3-brane with one
angular momentum was shown to reproduce the metric and the gauge
fields of the large ($k=0^+$ limit) of $D=5$ gauged supergravity black
holes \cite{Behrndt2}, with the anticipation that the result would
generalises to multiple angular momenta.  However, the identification
of the scalar fields was not given. (In addition, in
\cite{Cveticgubser1,Cveticgubser2}, the equivalence of the
thermodynamics of the near-extreme spinning branes and the
corresponding large black holes of $D=4,\ 5,\ 7$ gauged supergravity
was given.)  While incomplete, these works provided some initial
stages in the investigation of the sphere compactifications of
M-/string theory.

Unlike black holes that are asymptotically Minkowskian, for which the
horizons are always spherical, it is known that AdS black holes can
also admit horizons of more general topology.  Following the embedding
procedure described above, we demonstrate that AdS$_{4}$ black hole
with toroidal horizon can indeed be interpreted as the near-horizon
structures of an M2-brane rotating in the extra dimensions. The four
charges corresponding to the $U(1)^{4}$ Cartan subgroup are just the
four angular momenta. Similarly, the 5-dimensional charged black hole
with toroidal horizon corresponds to a rotating D3-brane and the
7-dimensional charged black hole with toroidal horizon to a rotating
M5-brane.  In each case, the event horizon coincides with the
worldvolume of the brane.\footnote{This is a concrete realisation of
the ``Membrane Paradigm'' \cite{thorne}.}
Additionally, one may use the Kaluza-Klein 
ansatz to obtain the higher-dimensional interpretation of AdS black holes
with horizons of other topologies.  We conjecture that these
correspond to the near-horizon limits of rotating $p$-branes whose
world-volumes have these topologies.  (In fact the rotating
``test'' membrane in \cite{BDPS2} had $S^2$ topology.)

In this paper we also obtain the general rotating
$p$-brane solutions in arbitrary dimensions, supporeted by a single
$(p+2)$-form charge, and discuss their sphere
reductions.  These rotating $p$-branes are easily constructed, merely
by performing standard diagonal dimensional oxidations of the general
rotating black holes that were constructed in \cite{Cveticyoum3}.

\section{$S^5$ reduction of type IIB supergravity}

        The $S^5$ reduction \cite{Schwarz,GM,KRV} of type IIB
supergravity gives rise to $N=8$, $D=5$ gauged supergravity, with
$SO(6)$ Yang-Mills gauge group \cite{PPV,GRW}.  The complete
details of this reduction, as with any sphere reduction, would be of
great complexity, and in fact no example has ever been fully worked
out.  For our present purposes, however, it suffices to consider the
truncation of the five-dimensional theory to $N=2$ supersymmetry.  In
this truncation, which is of course a consistent one, the gauge group
is reduced down to the $U(1)\times U(1)\times U(1)$ Cartan subgroup of
$SO(6)$.  The bosonic sector of the theory comprises these three gauge
bosons, the metric, and two scalar fields.  (The consistency of the
truncation to this field content can be seen by considering the $S^1$
reduction of ungauged minimal non-chiral supergravity in $D=6$, whose
bosonic fields $(g_{\mu\nu}, A_\2,\phi)$ reduce to give precisely the
field content we are considering here in $D=5$.  After gauging, one
would obtain the $U(1)^3$ gauged theory.)

\subsection{Reduction ans\"atze}

   Even to construct the $S^5$ reduction ansatz for this truncated
$N=2$ theory is somewhat non-trivial, owing to the presence of the
scalar fields.  It is most conveniently expressed in terms of the
parameterisation of sphere metrics given in \cite{mype}.

We find that the ansatz for the reduction of the
ten-dimensional metric is
\be
ds_{10}^2 = \sqrt{\wtd\Delta}\, ds_5^2 + \fft1{g^2\,
\sqrt{\wtd\Delta}}\,  \sum_{i=1}^3 X_i^{-1}\,
\Big(d\mu_i^2 + \mu_i^2\, (d\phi_i +g\, A^i)^2\Big)\ ,\label{2bs5met}
\ee
where the two scalars are parameterised in terms of the three
quantities $X_i$, which are subject to the constraint $X_1\, X_2\,
X_3=1$.  They can be parameterised in terms of two dilatons $\varphi_1$
and $\varphi_2$ as
\be
X_i=e^{-\ft12 \vec a_i\cdot \vec \varphi}\ ,
\ee
where $\vec a_i$ satisfy the dot products
\be
M_{ij}\equiv\vec a_i\cdot \vec a_j=4\delta_{ij} -\ft43\ .
\ee
A convenient choice is
\be
\vec a_1 = (\ft{2}{\sqrt6}, \sqrt2)\ , \qquad
\vec a_2 = (\ft{2}{\sqrt6}, -\sqrt2)\ ,\qquad
\vec a_3 = (-\ft{4}{\sqrt6}, 0)\ .
\ee
The three quantities $\mu_i$ are subject to the constraint $\sum_i
\mu_i^2=1$, and the metric on the unit round 5-sphere can be written
in terms of these as
\be
d\Omega_5^2 = \sum_i (d\mu_i^2 + \mu_i^2\, d\phi_i^2)\ .
\ee
The $\mu_i$ can be parameterised in terms of angles on a 2-sphere, for
example as
\be
\mu_1 = \sin\theta\ ,\qquad \mu_2 = \cos\theta\, \sin\psi\ ,\qquad
\mu_3 = \cos\theta\, \cos\psi\ .
\ee
Note that $\wtd\Delta$ is given by
\be
\wtd \Delta = \sum_{i=1}^3 X_i\, \mu_i^2\ ,
\ee
and is therefore expressed purely in terms of the scalar fields, and the
coordinates on the compactifying 5-sphere.  The constant $g$ in
(\ref{2bs5met}) is the inverse of the radius of the compactifying
5-sphere, and is equal to the gauge coupling constant.  We find that
the ansatz for
the reduction of the 5-form field strength is
$F_\5=G_\5+ {*G_\5}$, where
\bea
G_\5 &=& 2g\, \sum_i\Big(X_i^2\, \mu_i^2 -\wtd\Delta\, X_i\Big)\,
\ep_\5 - \fft1{2g}\, \sum_i X_i^{-1}\, {{\bar *} dX_i}\wedge
d(\mu_i^2) \nn\\
&&+ \fft1{2 g^2} \,
\sum_i X_i^{-2}\, d(\mu_i^2)\wedge (d\phi_i +g\, A_\1^i)
\wedge {{\bar *} F_\2^i}\ .\label{5fs5red}
\eea
Here, $F_\2^i=dA_\1^i$, $\ep_\5$ is the volume form of the 5-dimensional
metric $ds_5^2$, and ${\bar *}$ denotes the Hodge dual with respect to
the five-dimensional metric $ds_5^2$.

   Substituting these ans\"atze into the equations for motion for the
type IIB theory, we obtain five-dimensional
equations of motion that can be derived from the
Lagrangian\footnote{We shall make some more detailed comments on
certain general features of these spherical Kaluza-Klein reductions in
section 3, where we consider the $S^7$ reduction of $D=11$ supergravity.}
\be
e^{-1}\, {\cal L}_5 = R - \ft12(\del\varphi_1)^2 -\ft12(\del\varphi_2)^2
+ 4g^2\, \sum_i X_i^{-1}- \ft14 \sum_i X_i^{-2}\, (F_\2^i)^2  +\ft14
\ep^{\mu\nu\rho\sigma\lambda}\, F^1_{\mu\nu}\, F^2_{\rho\sigma}\,
A^3_\lambda\ .
\label{d5gauged}
\ee
(The other bosonic fields of the type IIB theory are set to zero in
this $U(1)^3$ truncated reduction.)  Note that the ten-dimensional
Bianchi identity $dF_\5=0$ gives rise to the equations of motion for
the scalars and gauge fields in five dimensions.

   Thus we have established that the reduction ans\"atze
(\ref{2bs5met}) and (\ref{5fs5red}) describe the exact embedding of
the five-dimensional $N=2$ gauged $U(1)^3$ supergravity into type IIB
supergravity.

        The bosonic Lagrangian (\ref{d5gauged}) can be further
truncated down to smaller sectors.  For example, we can consistently
set $\varphi_2=0$, implying that $X_1=X_2=X_3^{-1/2}$, provided that
$F_\2^1=F_\2^2=F_\2/\sqrt2$.  The Lagrangian then becomes
\bea
e^{-1} {\cal L}_5 &=& R-\ft12 (\del\varphi_1)^2 +
4g^2\, (2e^{\ft1{\sqrt6}\varphi_1} + e^{-\ft2{\sqrt6}\varphi_1}) -
\ft14e^{\ft2{\sqrt6}\varphi_1}\, (F_\2)^2-
\ft14e^{-\ft4{\sqrt6}\varphi_1}\, (F_\2^3)^2\nn\\
&& + \ft18 \ep^{\mu\nu\rho\sigma\lambda}\, F_{\mu\nu}\, F_{\rho\sigma}\,
A^3_\lambda\ .
\eea
It is also possible to set both scalars to zero, implying that
$X_i=1$, provided that $F_\2^i
=F_\2/\sqrt3$.  The Lagrangian is then given by
\be
e^{-1}{\cal L}_5 = R + 12 g^2 - \ft14 F_\2^2 +
\ft1{12\sqrt3} \ep^{\mu\nu\rho\sigma\lambda}\,
F_{\mu\nu}\, F_{\rho\sigma}\, A_\lambda\ .\label{d5trunc}
\ee
The embedding of the truncated Lagrangian (\ref{d5trunc}) in $D=10$
dimensions was discussed in \cite{Chamblin}.

\subsection{$D=5$ AdS black holes}

       The Lagrangian (\ref{d5gauged}) admits a three-charge AdS black
hole solution, given by \cite{Behrndt2}
\bea
ds_5^2 &=& -(H_1H_2H_3)^{-2/3}\, f\, dt^2 +
(H_1H_2H_3)^{1/3}\, (f^{-1}\, dr^2 + r^2 d\Omega_{3,k}^2)\ ,\nn\\
X_i&=& H_i^{-1}\, (H_1H_2H_3)^{1/3}\ ,\qquad
A^i_\1 = \sqrt{k}\, (1-H_i^{-1})\, \coth\beta_i\, dt\ ,
\label{d5adsbh}
\eea
and
\be
f=k-\fft{\mu}{r^2} + g^2\, r^2\, (H_1H_2H_3)\ ,\qquad
H_i = 1 + \fft{\mu\, \sinh^2\beta_i}{k\, r^2}\ .
\ee
Here $k$ can be 1, 0 or $-1$, corresponding to the foliating surfaces
of the transverse space being $S^3$, $T^3$ or $H^3$, with unit metric
$d\Omega_{3,k}^2$, where $H^3$ denotes the hyperbolic 3-space of
constant negative curvature.  In the case of $k=0$, one first needs to
make the rescaling \cite{Cveticgubser1} $\sinh^2\beta_i
\longrightarrow k\, \sinh^2\beta_i$, followed by sending $k$ to zero.
The gauge potential for $k=0$ case is then given by
\be
A^i_\1 =\fft{1-H_i^{-1}}{\sinh\beta_i}\, dt\ .
\ee

\subsection{Rotating D3-brane}

         In this section, we show that the $k=0$ three-charge AdS
black hole of the $N=2$ gauged supergravity in $D=5$ given in
(\ref{d5adsbh}) can be embedded in $D=10$ as a solution that is
precisely the decoupling limit of the rotating D3-brane.
The higher-dimensional solutions corresponding to five-dimensional AdS
black holes with $k=1$ and $k=-1$ can also be easily obtained, by
substituting the five-dimensional solutions into the $S^5$ reduction
ans\"atze.

          There can be three angular momenta, $\ell_i$, $i=1,2,3$, in
the rotating D3-brane.  The generic single-charge rotating $p$-branes,
which can be obtained by dimensional oxidation of the generic
single-charge rotating black holes constructed in \cite{Cveticyoum3},
are presented in appendix A.  We find that the metric of the rotating
D3-brane is given by%
\footnote{This metric agrees with previously obtained results
\cite{klt,rosf}, after correcting some typographical errors.}
%
\bea
ds_{10}^2 &=& H^{-\ft12}\Big(-(1 -\fft{2m}{r^4\Delta})
\, dt^2 + dx_1^2 + dx_2^2 + dx_3^2 \Big)
+ H^{\ft12}\Big[\fft{\Delta\, dr^2}{H_1H_2H_3 -2m\, r^{-4}}\nn\\
&&+ r^2\, \sum_{i=1}^3 H_i\,(d\mu_i^2 + \mu_i^2\, d\phi_i^2)
-\fft{4m\, \cosh\a}{r^4\, H\, \Delta}\, dt\, (\sum_{i=1}^3
\ell_i\, \mu_i^2\, d\phi_i) \nn\\
&& + \fft{2m}{r^4\, H\, \Delta}\, (\sum_{i=1}^3
\ell_i\, \mu_i^2\, d\phi_i)^2\Big]\ ,
\label{d3rotate}
\eea
where the functions $\Delta$, $H$, and $ H_i$ are given by
\bea
\Delta &=& H_1H_2H_3\, \sum_{i=1}^3 \fft{\mu_i^2}{H_i}\ , \qquad
H = 1 + \fft{2m\, \sinh^2\a}{r^4 \Delta}\ ,\nn\\
H_i &=& 1 + \fft{\ell_i^2}{r^2}\ ,\qquad i=1,2,3\ .
\label{d3rotatefun}
\eea
The rotating D3-brane is supported by the self-dual 5-form field
strength $F_\5$ of the type IIB theory.  It is given by $F_\5 = G_\5 +
*G_\5$, where $G_\5=dB_\4$ and
\be
B_\4 = \fft{1-H^{-1}}{\sinh\a}\Big(-\cosh\a\, dt + \sum_{i=1}^3 \ell_i\,
\mu_i^2\, d\phi_i\Big)\wedge d^3x\ .
\ee

       As is well known, the non-rotating D3-brane has a ``decoupling
limit'' where the spacetime of the D3-brane becomes a product space
$M_5\times S^5$.  If the D3-brane is extremal, $M_5$ is a
five-dimensional anti-de Sitter spacetime.  More generally, when the
D3-brane is non-extremal, $M_5$ is the Carter-Novotny-Horsky metric
\cite{clp1}, which can thus be viewed as a ``non-extremal''
generalisation of AdS$_5$.  A similar limit also exists for the
rotating D3-brane, and can be achieved by making the rescalings
\bea
&&m\longrightarrow \ep^4 \, m\ ,\qquad
\sinh\a \longrightarrow \ep^{-2}\, \sinh\a\ ,\nn\\
&&r\longrightarrow \ep\, r\ ,
\qquad x^\mu\longrightarrow \ep^{-1}\, x^\mu\ ,
\qquad \ell_i \rightarrow \ep\, \ell_i\ ,
\eea
and then sending $\ep\longrightarrow 0$.  (Note that when this limit
is taken, we also have $\cosh\a\longrightarrow \ep^{-2}\, \sinh\a$.)
This has the effect that the last term in (\ref{d3rotate}) is set to
zero and that
\be
H = 1 + \fft{2m\, \sinh^2\a}{r^4 \Delta} \longrightarrow
\fft{2m\, \sinh^2\a}{r^4 \Delta}\ .
\ee
In this limit, the metric (\ref{d3rotate}) becomes
\bea
ds_{10}^2 &=& \sqrt{\wtd\Delta}\, \Big[-(H_1 H_2 H_3)^{-2/3}\,
f\, dt^2 + (H_1 H_2 H_3)^{1/3}(f^{-1}\, dr^2 + r^2\,
d\vec y\cdot d\vec y)\Big]\nn\\
&&+\fft1{g^2\,\sqrt{\wtd \Delta}}\, \sum_{i=1}^3 X_i^{-1}\,
\Big(d\mu_i^2 + \mu_i^2\, (d\phi_i +g\, A^i)^2\Big)
\ ,\label{d3rotatehor}
\eea
where
\be
\vec y = g\, \vec x\ ,\qquad
g^2 =\fft{1}{\sqrt{2m}\,\sinh\a}\ ,\qquad
\mu = 2m\,g^2\ .
\ee
The metric (\ref{d3rotatehor}) precisely matches the dimensional
reduction ansatz (\ref{2bs5met}), with the lower dimensional fields
given by
\bea
ds_5^2 &=& -(H_1 H_2 H_3)^{-2/3}\,
f\, dt^2 + (H_1\, H_2\, H_3)^{1/3}(f^{-1}\, dr^2 + r^2\,
d\vec y\cdot d\vec y)\ ,\nn\\
X_i &=& H_i^{-1}\, (H_1 H_2 H_3)^{1/3}\ , \qquad
A^i_\1 = \fft{1 - H_i^{-1}}{g\,\ell_i\,\sinh\a}\, dt\ ,
\label{d5adsbhk0}
\eea
where
\be
f=-\fft{\mu}{r^2} + g^2\, r^2\, H_1 H_2 H_3\ ,\qquad
g^2=\fft{1}{\sqrt{2m}\sinh\a}\ ,\qquad \mu = 2m\, g^2\ .
\ee
         To complete the story, we note that the 5-form field strength
in the decoupling limit is given by $F_\5=G_\5+{*G_\5}$, where
$G_\5=dB_\4$ and
\be
B_\4 = -g^4\, r^4\, \Delta\, dt\wedge d^3x + \fft{1}{\sinh\a}\,
(\sum_{i=1}^3 \ell_i\, \mu_i^2\, d\phi_i)\wedge d^3x\ .
\ee
This gives precisely the field strength in the
dimensional reduction ansatz (\ref{5fs5red}).

   Thus we see that the solution (\ref{d5adsbhk0}) is precisely the
$k=0$ three-charge AdS black hole given in the previous subsection,
after reparameterising the angular momenta $\ell_i^2 = \mu
\sinh^2\beta_i$.  This shows that the embedding of the three-charge
AdS $k=0$ black hole in gauged $N=2$ supergravity in five dimensions
gives a ten-dimensional solution that is precisely the decoupling
limit of the rotating D3-brane.  Single-charge AdS black holes coming
from the reduction of the metric of a rotating D3-brane with one
angular momentum was obtained in \cite{Cveticgubser1}, however
without the explicit embedding of the scalar fields.  The connection
between the thermodynamics of AdS black holes and rotating $p$-branes
was discussed in \cite{Cveticgubser1,Cveticgubser2}.

         It is also straightforward to oxidise the $k=1$ and $k=-1$
AdS black holes back to $D=10$ type IIB.  The metric is the same form
as (\ref{d3rotatehor}) with $d\vec{y}\cdot d\vec y$ replaced by the
unit metric for $S^3$ or $H^3$ respectively.  The 5-form field
strength follows by substituting the five-dimensional fields into
(\ref{5fs5red}).

\section{$S^7$ reduction of $D=11$ supergravity}

\subsection{Reduction ans\"atze}

   The $S^7$ reduction of eleven-dimensional supergravity gives rise
to $SO(8)$ gauged $N=8$ supergravity in four dimensions.  One may
again consider a consistent truncation to $N=2$, for which the bosonic
sector comprises the metric, four commuting $U(1)$ gauge potentials,
three dilatons and three axions.  (That this is a consistent
truncation can be seen by reducing minimal non-chiral six-dimensional
supergravity on $T^2$, for which the reduction of $(g_{\mu\nu}, A_\2,
\phi)$ will give precisely the field content we are considering.
After gauging, this would give the $U(1)^4$ gauged theory.  See
appendix B for an extended discussion of this.) We have not yet
determined the complete reduction ansatz for the entire truncated
theory where the axions are included, but we can give the exact ansatz
in the case where one sets the axions to zero.  This will not, of
course, be a consistent truncation, since the $U(1)$ gauge fields will
provide source terms of the form $\ep^{\mu\nu\rho\sigma}\,
F_{\mu\nu}\, F_{\rho\sigma}$ for the axions.  Nevertheless, one can
use the axion-free ansatz for discussing the exact embedding of
four-dimensional solutions for which the axions are zero.  The full
$N=2$ four-dimensional theory, including the axions, is obtained in
appendix B.

   The reduction ansatz for the eleven-dimensional metric is
\be
ds_{11}^2 = \wtd\Delta^{2/3}\, ds_4^2 +g^{-2}\,
\wtd\Delta^{-1/3}\, \sum_i X_i^{-1}\, \Big( d\mu_i^2 + \mu_i^2\,
 (d\phi_i + g\,
A^i_\1)^2 \Big)\ .\label{s7metred}
\ee
where $\wtd \Delta = \sum_{i=1}^4 X_i\, \mu_i^2$.  The four quantities
$\mu_i$ satisfy $\sum_i \mu_i^2 =1$.  They can be parameterised in
terms of angles on the 3-sphere as
\be
\mu_1 =\sin\theta\ ,\quad \mu_2=\cos\theta\, \sin\varphi\ ,\quad
\mu_3=\cos\theta\, \cos\varphi\, \sin\psi\ ,\quad
\mu_4=\cos\theta\, \cos\varphi\, \cos\psi\ .
\ee
The four $X_i$, which satisfy $X_1X_2X_3X_4=1$, can be parameterised
in terms of three dilatonic scalars $\vec\varphi =(\varphi_1,
\varphi_2, \varphi_3)$:
\be
X_i=e^{-\ft12 \vec a_i \cdot \vec \varphi}\ ,
\ee
where the $\vec a_i$ satisfy the dot products
\be
M_{ij} \equiv \vec a_i\cdot \vec a_j = 4\delta_{ij} -1\ .
\ee
A convenient choice, corresponding to the combinations of
(\ref{eq:lambdas}), is
\be
\vec a_1 =(1,1,1)\ ,\quad \vec a_2=(1,-1,-1)\ ,\quad
\vec a_3=(-1,1,-1)\ ,\quad \vec a_4=(-1,-1,1)\ .
\ee

   The reduction ansatz for the 4-form field strength is
\bea
F_\4 &=& 2g\,\sum_i \Big(X_i^2\, \mu_i^2 - \wtd\Delta\, X_i \Big)\,
\ep_\4 +\fft1{2g}\, \sum_i X_i^{-1}\, {{\bar *}dX_i}\wedge d(\mu_i^2)
\nn\\
&&-\fft1{2g^2}\, \sum_i X_i^{-2}\, d(\mu_i^2)\wedge (d\phi_i + g\,
A^i_\1) \wedge
{{\bar *} F_\2^i}\ .\label{s7f4red}
\eea
Here, ${\bar *}$ denotes the Hodge dual with respect to the
four-dimensional metric $ds_4^2$, and $\ep_\4$ denotes its
 volume form.\footnote{If the ans\"atze (\ref{s7metred}) and
(\ref{s7f4red}) are linearised around an AdS$_4\times S^7$ background,
they can be seen to be in agreement with previous results that were
derived at the linear level \cite{duffpope3}.  The full non-linear
metric ansatz (\ref{s7metred}) should be in agreement with the
appropriate specialisation of the ansatz
given in \cite{nilsson}.}

         It is of interest to note that the eleven-dimensional Bianchi
identity $dF_\4=0$ already gives rise to the four-dimensional
equations of motion for the scalars and gauge potentials, namely
\bea
d{{\bar *}d\log(X_i)} &=& \ft14 \sum_{j} M_{ij}\, X_j^{-2}\,
{{\bar *}F_\2^j}\wedge F_\2^j +g^2\sum_{j,k} M_{ij}\, X_j\, X_k
-g^2\sum_j M_{ij} X_j^2\ ,\nn\\
d(X_i^{-2}\, {{\bar *}F_\2^i}) &=&0\ .
\eea
It is straightforward to see that these equations of motion can be
obtained from the four-dimensional Lagrangian
\be
e^{-1}{\cal L}_4 = R -
\ft12 (\del\vec\varphi)^2 +
8g^2 (\cosh\varphi_1 +\cosh\varphi_2+\cosh\varphi_3)
-\ft14 \sum_{i=1}^4 e^{\vec a_i\cdot\vec \varphi}\,
(F_\2^i)^2\ .\label{d4lagxx}
\ee

    One might think that it would be possible to obtain the
four-dimensional Lagrangian by substituting the ans\"atze
(\ref{s7metred}) and (\ref{s7f4red}) into the eleven-dimensional
Lagrangian.  In fact this is not the case, and one must work at the
level of the eleven-dimensional equations of motion.  One way of
understanding this is from the fact that the ansatz for $F_\4$ does
not {\sl identically} satisfy the Bianchi identity.  Rather, as we
have seen, it satisfies it modulo the use of the four-dimensional
equations of motion for the scalars and gauge fields.  In other words,
the ansatz is made on the eleven-dimensional 4-form $F_\4$ rather than
on the fundamental potential $A_\3$ itself.  Consequently, it would
not be correct to insert the ansatz for $F_\4$ into the Lagrangian.

     We may further illustrate this point by showing, as an example,
how the scalar potential arises in the four-dimensional Einstein
equation.  This comes from considering the eleven-dimensional Einstein
equation,
\be
\hat R_{AB} - \ft12 \hat R\, \eta_{AB} = \ft1{12} \Big( F^2_{AB}
- \ft18 F^2\, \eta_{AB}\Big)\ .\label{d11einst}
\ee
with vielbein indices $A,B$ ranging just over the four-dimensional spacetime
directions $\a,\b$.  From the ansatz (\ref{s7metred}), the relevant terms in
the eleven-dimensional Ricci tensor and Ricci scalar are given by
\bea
\hat R_{\a\b}&=&
\fft{4g^2}{3\wtd\Delta^{8/3}}\, \Big[ -\Big( \sum_i X_i^2\,
\mu_i^2\Big)^2 +\wtd \Delta\, \sum_i X_i^2\, \mu_i^2\, \sum_j X_j +
\wtd\Delta\, \sum_i X_i^3\, \mu_i^2 \nn\\
&&\qquad  - \wtd\Delta^2\, \sum_i
X_i^2 \Big]\, \eta_{\a\b} + \wtd \Delta^{-2/3}\, \, R_{\a\b} +
\cdots\ ,\label{ricci2}\\
\hat R &=& \fft{2g^2}{3\wtd\Delta^{8/3}}\, \Big[ -\Big( \sum_i X_i^2\,
\mu_i^2\Big)^2 - 2\wtd \Delta\, \sum_i X_i^2\, \mu_i^2\, \sum_j X_j +
4 \wtd\Delta\, \sum_i X_i^3\, \mu_i^2 \nn\\
&&\qquad + 6\wtd\Delta^2\, \Big(\sum_i X_i\Big)^2 - 7\wtd\Delta^2\, \sum_i
X_i^2 \Big] + \wtd\Delta^{-2/3}\, R +\cdots\ ,\label{ricci1}
\eea
where $R_{\a\b}$ and $R$ are the four-dimensional Ricci tensor and
scalar, and the ellipses indicate that terms not involving purely the
undifferentiated scalars have been omitted for the purposes of the
present illustrative discussion.  From the ansatz
(\ref{s7f4red}) for the 4-form, the eleven-dimensional energy-momentum
tensor vielbein components in the four-dimensional spacetime
directions are given by
\be
\ft1{12}(F^2_{\a\b} - \ft18 F^2\, \eta_{a\b}) =
-g^2\, \wtd\Delta^{-8/3}\, \Big(\sum_i(X_i^2\, \mu_i^2 -\wtd\Delta\,
X_i)\Big)^2 \, \eta_{\a\b}\ .\label{enmom}
\ee

   Substituting (\ref{ricci2}), (\ref{ricci1}) and (\ref{enmom}) into
(\ref{d11einst}), we find that all the angular dependence coming from
the $\mu_i$ variables cancels, and that the scalar
potential terms in the four-dimensional Einstein equation are given by
\be
R_{\a\b} -\ft12 R\, \eta_{\a\b} = -\ft12 g^2\, V\, \eta_{\a\b}\ ,
\label{einpot}
\ee
with $V$ given by
\be
V= - 4 \sum_{i<j} X_i\, X_j = -8 (\cosh\varphi_1 + \cosh\varphi_2 +
\cosh\varphi_3)\ .
\ee
Since (\ref{einpot}) derives from the Lagrangian $R -g^2\, V$, we see
that we have precisely produced the hoped-for potential terms of the
gauged supergravity Lagrangian (\ref{d4lagxx}).  This sample
calculation also serves to illustrate that the angular dependence
coming from the $\mu_i$ variables would not have cancelled if we had
merely substituted the ans\"atze (\ref{s7metred}) and (\ref{s7f4red})
into the eleven-dimensional Lagrangian.  It also shows that the
cancellation of the $\mu_i$ dependence in the higher-dimensional
equations of motion depends crucially on ``conspiracies'' between the
contributions from the metric and the 4-form field strength.  This is
quite different from the situation in toroidal reductions, where each
term in the higher-dimensional theory reduces consistently by itself,
without the need for any such conspiracies.  Note, furthermore, that
the required conspiracies needed for the success of the spherical
reduction depend on the 4-form field strength occurring with precisely
the correct coefficient relative to the Einstein-Hilbert term.
This normalisation is not a free parameter, but is governed by the
strength of the $FFA$ term in the eleven-dimensional theory.  Thus
ultimately the consistency of the spherical reduction ansatz can be
traced back to the supersymmetry of the eleven-dimensional theory.

    Note that the Lagrangian (\ref{d4lagxx}) can be further truncated,
to pure Einstein-Maxwell with a cosmological constant, by setting all
the field strengths equal, $F_\2^i = \ft12 F_\2$, and setting all the
scalars to zero:
\be
e^{-1}{\cal L}_{4} = R -\ft14 (F_\2)^2 + 24 g^2\ .
\ee
The embedding of this theory into $D=11$ supergravity was obtained in
\cite{Pope}. The ansatz for the metric and field strength for the
embedding in \cite{Pope} was given in terms of a decomposition of the
7-sphere as a $U(1)$ bundle over $CP^3$.  This is identical, after a
transformation of coordinates, to the Einstein-Maxwell embedding given
in \cite{Chamblin}.  In the same spirit, the $S^5$ reduction to
five-dimensional Einstein-Maxwell can be described using the method
presented in \cite{Pope}, with $S^5$ viewed as a $U(1)$ bundle over
$CP^2$.  (An analogous consistent embedding of four-dimensional
Einstein-Yang-Mills with an $SU(2)$ gauge group, and a cosmological
constant, in $D=11$ supergravity was obtained in \cite{Pope2}. This
involves a decomposition of $S^7$ as an $SU(2)$ bundle over $S^4$.)

\subsection{$D=4$ AdS black holes}

    The $D=4$, $N=2$ gauged supergravity coupled to three vector
multiplets  admits 4-charge AdS black hole solutions, given by
\cite{Duffliu,Sabra}
\bea
ds_4^2 &=& -(H_1H_2H_3H_4)^{-1/2}\, f\, dt^2 +
(H_1H_2H_3)^{1/2}\, (f^{-1}\, dr^2 + r^2 d\Omega_{2,k}^2)\ ,\nn\\
X_i&=& H_i^{-1}\, (H_1H_2H_3H_4)^{1/4}\ ,\qquad
A^i_\1 = \sqrt{k}\, (1-H_i^{-1})\, \coth\beta_i\, dt\ ,
\label{d4adsbh}
\eea
and
\be
f=k-\fft{\mu}{r} + 4g^2\, r^2\, (H_1H_2H_3H_4)\ ,\qquad
H_i = 1 + \fft{\mu\, \sinh^2\beta_i}{k\, r}\ .\label{d4adsbh2}
\ee
Here, $k$ can be 1, 0 or $-1$, corresponding to the cases where the
foliations in the transverse space have the metric $d\Omega_{2,k}^2$
on the unit $S^2$, $T^2$ or $H^2$, where $H^2$ is the unit hyperbolic
2-space of constant negative curvature.
In the case of $k=0$, one must first make the
rescaling $\sinh^2\beta_i \longrightarrow k\, \sinh^2\beta_i$
before sending $k$ to zero.  The gauge potential for the $k=0$ case is
then given by
\be
A^i_\1 =\fft{1-H_i^{-1}}{\sinh\beta_i}\, dt\ .
\ee

\subsection{Rotating M2-brane}

      There are four angular momenta, $\ell_i$, $i=1,2,3,4$, in the
rotating M2-brane.  The solution can be obtained by oxidising the
$D=9$ rotating black hole \cite{Cveticyoum3}.  After the oxidation, we
find that the metric of the rotating M2-brane is given by
\bea
ds_{11}^2&=& H^{-\ft23}\, \Big(-(1 -\fft{2m}{r^6\, \Delta})\, dt^2 +
dx_1^2 + dx_2^2 \Big) + H^{\ft13}\Big[
\fft{\Delta\, dr^2}{H_1\, H_2\, H_3\, H_4 -\fft{2m}{r^6}}\nn\\
&&+ r^2 \sum_{i=1}^4 H_i\, (d\mu_i^2 + \mu_i^2\, d\phi_i^2) -
\fft{4m\, \cosh\a}{r^6\, H\, \Delta}\, dt
(\sum_{i=1}^4 \ell_i\, \mu_i^2\, d\phi_i)\nn\\
&&+\fft{2m}{r^4\, H\, \Delta}\, (\sum_{i=1}^4 \ell_i\,
\mu_i^2\, d\phi_i)^2 \Big]\ ,\label{m2rotate}
\eea
where the functions $\Delta$, $H$ and $H_i$ are given by
\bea
\Delta&=& H_1 H_2 H_3 H_4\, \sum_{i=1}^4 \fft{\mu_i^2}{H_i}
\ ,\qquad H=1 +\fft{2m\, \sinh^2\a}{r^6\, \Delta}\ ,\nn\\
H_i&=& 1 + \fft{\ell_i^2}{r^2}\ ,\qquad i=1,2,3,4\ .\label{funct2}
\eea
The 3-form gauge potential is given by
\be
A_\3 = \fft{1-H^{-1}}{\sinh\a} (-\cosh\a\, dt +
\ell_i\, \mu_i^2\, d\phi_i)\wedge d^2x\ .\label{a3form}
\ee

        Following the previous D3-brane example, we consider the
decoupling limit, which is obtained by making the rescaling
\bea
&&m\longrightarrow \ep^6 m\ ,\qquad
\sinh\a \longrightarrow \ep^{-3}\, \sinh\a\ ,\nn\\
&&r\longrightarrow \ep\, r\ ,
\qquad x^\mu\longrightarrow \ep^{-2}\, x^\mu\ ,
\qquad \ell_i \rightarrow \ep\, \ell_i
\eea
and then sending $\ep\longrightarrow 0$.  This has the effect that the
last term in (\ref{m2rotate}) is set to zero and that the 1 in
function $H$ (\ref{funct2}) is removed.   In this limit,
the rotating M2-brane (\ref{m2rotate}) becomes
\bea
ds_{11}^2 &=& \wtd\Delta^{2/3}\, \Big[ -(H_1 H_2 H_3 H_4)^{-1/2}\, f\,
dt^2 + (H_1 H_2H_3H_4)^{1/2}\, (f^{-1}\, d\rho^2 + \rho^2 d\vec y
\cdot d\vec y)\Big] \nn\\
&&+ g^{-2}\, \wtd\Delta^{-1/3}\, \sum_{i=1}^4 X_i^{-1}\Big(d\mu_i^2 +
\mu_i^2\, (d\phi_i + g\, A^i)^2\Big)\ ,\label{m2limmet}
\eea
where
\bea
&&\rho= \ft12 g\, r^2\ , \qquad \vec y= 2g\, \vec x\ ,\qquad
f=-\fft{\mu}{\rho} + 4g^2\, \rho^2\, H_1 H_2 H_3 H_4\ ,\nn\\
&&g^2=(2m\,\sinh^2\a)^{-1/3}\ ,\qquad \mu = m\, g^5\ ,\qquad
\wtd\Delta = \sum_i\, X_i\, \mu_i^2 \ .
\eea
This is precisely of the form of the metric ansatz in the dimensional
reduction given by (\ref{s7metred}).  The lower dimensional fields are
given by
\bea
&&ds^2_4 = -(H_1 H_2 H_3 H_4)^{-1/2}\, f\,
dt^2 + (H_1 H_2H_3H_4)^{1/2}\, (f^{-1}\, d\rho^2 + \rho^2 d\vec {\td
x}\cdot d\vec {\td x})\nn\\
&&X_i = H_i^{-1}\, (H_1 H_2 H_3 H_4)^{1/4}\ , \qquad
A^i = \fft{1 - H_i^{-1}}{g\,\ell_i\,\sinh\a}\, dt\ .
\eea

   In the decoupling limit, the gauge potential $A_\3$ given in
(\ref{a3form}) for the rotating M2-brane becomes, after a gauge
transformation,
\be
A_\3 = -g^6\, r^6\, \Delta\, dt\wedge d^2x + \fft1{\sinh\a}\,
\sum_i \ell_i\, \mu_i^2\, d\phi_i\wedge d^2x\ .
\ee
We find that its field strength $F_\4=dA_\3$ is also of the form given
in (\ref{s7f4red}) for the dimensional reduction ansatz.  Thus we have
established an exact embedding of four-dimensional non-extremal
4-charge AdS black holes into eleven-dimensional supergravity, and
furthermore, that they become precisely the decoupling limit of the
rotating M2-branes.  It should, of course, be emphasised that the
four-dimensional AdS black holes that we are considering at this point
have $T^2$ rather than $S^2$ horizons, corresponding to $k=0$ in
(\ref{d4adsbh}) and (\ref{d4adsbh2}).

        It is also straightforward to oxidise the $k=1$ and $k=-1$ AdS
black hole solutions back to $D=11$, by substituting the
four-dimensional fields into the ans\"atze (\ref{s7metred}) and
(\ref{s7f4red}).

\section{$S^4$ reduction of $D=11$ supergravity}

\subsection{Reduction ans\"atze}

   The Kaluza-Klein reduction of eleven-dimensional supergravity on
$S^4$ gives rise to $N=4$ gauged $SO(5)$ supergravity in seven
dimensions.  In a similar manner to the $S^5$ and $S^7$ reductions
that we discussed previously, we may consider an $N=2$ truncation of
this seven-dimensional theory.  As described in the introduction, the
truncated theory comprises $N=2$ supergravity coupled to a vector
multiplet, comprising the metric, 2-form potential, four vector
potentials and four scalars in total.  For our present purposes, we we
shall focus on a further truncation where only the metric, two gauge
potentials (which are associated with the $U(1)\times U(1)$ Cartan
subgroup of $SO(5)$) and two scalars are retained.  This is not in general a
consistent truncation, but, as in the case of the neglect of the
axions in the $S^7$ reduction, it is consistent for a subset of
solutions where the truncated fields are not excited by the ones that
are retained.   In particular, solutions of the $N=2$ theory for which
$F_\2^1\wedge F_\2^2=0$, such as the AdS black holes, will also be
solutions of this truncated theory.

    We find that we can obtain this truncated theory by making the
following Kaluza-Klein $S^4$-reduction ansatz:
\bea
ds_{11}^2 &=& \wtd\Delta^{1/3}\, ds_7^2 + g^{-2}\, \wtd\Delta^{-2/3}\,
\Big(X_0^{-1}\, d\mu_0^2 + \sum_{i=1}^2 X_i^{-1}\, (d\mu_i^2 + \mu_i^2\,
(d\phi_i + g\, A_\1^i)^2) \Big)\ ,\label{s4metred}\\
{*F_\4} &=&
2g\,\sum_{\a=0}^2 \Big(X_\a^2\, \mu_\a^2 - \wtd\Delta\, X_\a \Big)\,
\ep_\7 + g\, \wtd\Delta\, X_0\, \ep_\7
+\fft1{2g}\, \sum_{\a=0}^2 X_\a^{-1}\, {{\bar *}dX_\a}
\wedge d(\mu_\a^2) \nn\\
&&+\fft1{2g^2}\, \sum_{i=1}^2 X_i^{-2}\, d(\mu_i^2)\wedge
(d\phi_i + g\, A^i_\1) \wedge
{{\bar *} F_\2^i}\ ,\label{s4f4red}
\eea
where we have defined the auxiliary variable $X_0\equiv (X_1 X_2)^{-2}$.
Here, ${\bar *}$ denotes the Hodge dual
with respect to the seven-dimensional metric $ds_7^2$, $\ep_\7$
denotes its volume form, and $*$ denotes the Hodge dualisation in the
eleven-dimensional metric.  The quantity $\wtd\Delta$ is given by
\be
\wtd\Delta = \sum_{\a=0}^2 X_\a\, \mu_\a^2\ ,
\ee
where $\mu_0$, $\mu_1$ and $\mu_2$ satisfy
$\mu_0^2+\mu_1^2+\mu_2^2=1$.   The two scalar fields $X_i$ can be
parameterised in terms of two canonically-normalised dilatons
$\vec\varphi=(\varphi_1,\varphi_2)$ by writing
\be
X_i = e^{-\fft12\vec a_i\cdot\vec\varphi}\ ,
\ee
where the dilaton vectors satisfy the relations $\vec a_i\cdot\vec a_j
= 4\delta_{ij} -\ft85$.  A convenient parameterisation is given by
\be
\vec a_1 = (\sqrt2, \sqrt{\ft25})\ ,\qquad \vec a_2 = (-\sqrt2,
\sqrt{\ft25}) \ .\label{d7adef}
\ee
Note that the two $X_i$ are independent here, unlike in the cases of
the three $X_i$ in $D=5$ or the four $X_i$ in $D=4$, which satisfied
$\prod_i X_i =1$.  The auxiliary variable $X_0$ that we have
introduced in order to make the expressions more symmetrical can be
written as $X_0=e^{-\fft12\vec a_0\cdot\vec\varphi}$, where $\vec a_0=
-2(\vec a_1+\vec a_2)= (0,-4\sqrt{2/5})$.

    After substituting into the eleven-dimensional equations of
motion, one obtains seven-dimensional equations that can be derived
from the Lagrangian
\be
e^{-1}{\cal L}_7 = R -\ft12 (\del\vec\varphi)^2 - g^2\, V -
\ft14 \sum_{i=1}^2 e^{\vec a_i\cdot\vec\varphi}\, (F_\2^i)^2\ ,
\label{d7lag}
\ee
where the potential $V$ is given by
\be
V= -4 X_1 X_2 - 2X_1^{-1}\, X_2^{-2} - 2 X_2^{-1}\, X_1^{-2} +\ft12
(X_1 X_2)^{-4}\ .
\ee
This potential has a more complicated structure than those in the
$D=5$ and $D=4$ gauged theories, and in particular it has not only a
maximum at $X_1=X_2=1$, but also a saddle point at $X_1=X_2=
2^{-1/5}$ \cite{ppv}.  Note that by making use of the auxiliary variable
$X_0=(X_1 X_2)^{-2}$, the potential can be re-expressed as
\be
V = -4 X_1 X_2 -2 X_0 X_1 -2 X_0 X_2 + \ft12 X_0^2\ .
\ee

   It is interesting to note that the Lagrangian (\ref{d7lag}) can be
further consistently truncated, by setting $X_1=X_2=X$, and $F_\2^1 =
F_\2^2=F_\2/\sqrt2$.  This implies that the dilatonic scalar
$\varphi_1$ is set to
zero, in terms of the parameterisation defined by (\ref{d7adef}).
This gives
\be
e^{-1}{\cal L}_7 = R - \ft12(\del\varphi_2)^2 + g^2\, ( 4X^2 +
4X^{-3} -\ft12 X^{-8}) -\ft14 X^{-2}\, (F_\2)^2\ ,
\ee
where
\be
X = e^{-\fft1{\sqrt{10}}\varphi_2}\ .
\ee
This scalar potential was used in \cite{lpss} to construct
supersymmetric domain
wall solutions.

\subsection{$D=7$ AdS black holes}
\la{seven}

    This Lagrangian (\ref{d7lag}) admits 2-charge AdS black-hole
solutions, given by
\bea
ds_7^2 &=& -(H_1H_2)^{-4/5}\, f\, dt^2 +
(H_1H_2)^{1/5}\, (f^{-1}\, d r^2 + r^2\, d\Omega_{5,k}^2)\ ,\nn\\
f&=& k -\fft{\mu}{r^4} + \ft14 g^2\, r^2\, H_1 H_2\ ,\qquad
X_i= (H_1H_2)^{2/5}\, H_i^{-1}\ ,\nn\\
A_\1^i &=&\sqrt k \, \coth\beta_i\, (1-H_i^{-1})\, dt\ ,\qquad
H_i = 1+ \fft{\mu\, \sinh^2\beta_i}{r^4}\ ,
\eea
where $d\Omega_{5,k}^2$ is the metric on a unit $S^5$, $T^5$ or $H^5$
according to whether $k=1,0$ or $-1$.  As in the previous cases we
discussed, the $k=0$ solution is obtained by first rescaling
$\sinh^2\beta_i \longrightarrow k\, \sinh^2\beta_i$ before setting
$k=0$.  The metric of the $D=7$ AdS black hole was
obtained in \cite{Cveticgubser1}, by isolating the spacetime
direction of the rotating M5-brane metric.

\subsection{Rotating M5-brane}

      There are two angular momenta, $\ell_1$ and $\ell_2$, in the
rotating M5-brane \cite{Cveticyoum2,Csaki2}.  Its metric is given by
\bea
ds_{11}^2 &=& H^{-1/3}\Big(-(1-\fft{2m}{r^3\Delta})\, dt^2 +
dx_1^2 + \cdots + dx_5^2\Big) +
H^{2/3}\Big[\fft{\Delta\, dr^2}{H_1H_2 - \fft{2m}{r^3}} \nn\\
&&+r^2\Big(d\mu_0^2 +\sum_{i=1}^2 H_i(d\mu_i^2 + \mu_i^2\,
d\phi_i^2)\Big) - \fft{4m\, \cosh\a}{r^3\, H\, \Delta}\, dt\,
(\sum_{i=1}^2 \ell_i\, \mu_i^2\, d\phi_i)^2\nn\\
&&+\fft{2m}{r^3\, H\, \Delta}(\sum_{i=1}^2\ell_i\, \mu_i^2\,
d\phi_i)^2\Big]\ ,
\eea
where $\Delta$, $H$ and $H_i$ are given by
\bea
&&\Delta = H_1 H_2(\mu_0^2 + \fft{\mu_1^2}{H_1} + \fft{\mu_2^2}{H_2})\ ,
\qquad H= 1 + \fft{2m\, \sinh^2\a}{r^3\, \Delta}\ ,\nn\\
&&H_1=1 + \fft{\ell_1^2}{r^2}\ ,\qquad
H_2=1 + \fft{\ell_2^2}{r^2}\ .
\eea
The three quantities $\mu_0$, $\mu_1$ and $\mu_2$ satisfy
$\mu_0^2+\mu_1^2 +\mu_2^2=1$.
The 4-form field strength is given by $F_4={*dA_6}$, where
\be
A_6 = \fft{1-H^{-1}}{\sinh\a}(\cosh\a\, dt +
\ell_1\, \mu_1^2\, d\phi_1+ \ell_2\, \mu_2^2\, d\phi_2)\wedge
d^5x\ .
\ee

   The decoupling limit is defined by
\bea
&&m\longrightarrow \ep^3\, m\ ,\qquad
\sinh\a \longrightarrow \ep^{-3/2}\, \sinh\a\ ,\nn\\
&&r\longrightarrow \ep\, r\ ,\qquad
x^{\mu} \longrightarrow \ep^{-1/2}\, x^\mu\ ,\qquad
\ell_i \longrightarrow \ep\, \ell_i\ ,
\eea
with $\ep \longrightarrow 0$.  In this limit, the metric becomes
\bea
ds_{11}^2&=& \wtd \Delta^{1/3}\Big[-(H_1H_2)^{-4/5}\, f\, dt^2 +
(H_1H_2)^{1/5}\, (f^{-1}\, d\rho^2 + \rho^2\, d\vec y\cdot d\vec y)
\Big]\nn\\
&&+g^{-2}\, \wtd \Delta^{-2/3}\,
\Big((X_1X_2)^{2}\, d\mu_0^2 +
\sum_{i=1}^2 X_i^{-1}\, (d\mu_i^2 + \mu_i^2(d\phi_i +
      g\, A_\1^i)^2 )\Big)
\ .\label{m5rotatehor}
\eea
where
\bea
&&\rho^2 =4r\, g^{-1}\ ,\qquad \vec y= \ft12g\, \vec x\ ,\qquad
\wtd \Delta=(X_1 X_2)^{-2}\, \mu_0^2 +
X_1\, \mu_1^2 + X_2\, \mu_2^2 \ ,\nn\\
&&g^2 =(2m\,\sinh^2\a)^{-2/3}\ ,\qquad \mu = 32 m\, g^{-1}\ .
\eea
The metric (\ref{m5rotatehor}) fits precisely the dimensional
reduction ansatz given in (\ref{s4metred}). The lower dimensional
fields are given by
\bea
ds_7^2&=&-(H_1H_2)^{-4/5}\, f\, dt^2 + (H_1H_2)^{1/5}\,
(f^{-1}\, d\rho^2 + \rho^2\, d\vec y\cdot d\vec y)\ ,\nn\\
X_i&=& (H_1 H_2)^{2/5}\, H_i^{-1}\ ,\qquad
 f=-\fft{\mu}{\rho^4} + \ft14 g^2\rho^2\, H_1H_2\ ,\nn\\
A_\1^i &=& \fft{1-H_i^{-1}}{g\, \ell_i\, \sinh\a}\, dt\ .
\eea
This is precisely the $k=0$ AdS$_7$ black hole obtained in previous
section, with the angular momenta reparameterised as $\ell_i=\mu\,
g^2\, \sinh^2\beta_i/16$.  This establishes that the 2-charge $k=0$
AdS black hole in $D=7$ can be reinterpreted as the decoupling limit
of the rotating M2-brane.  (Of course in this example, one can {\it
only} discuss the embedding when the scalar fields are included, since
there is no choice of charge parameters for which the scalar fields
vanish in the seven-dimensional black holes. This contrasts with the
cases of the rotating D3-branes and M5-branes, where the special
choice of setting all the charges equal allows the discussion of a
simplified ansatz where the scalars are omitted.)

\section{Sphere reduction of generic rotating $p$-branes, and domain
wall black holes}

    The general expression for a rotating $p$-brane carrying a single
charge is given in appendix A.  Following the procedure in the
previous sections, we may take the limit of large $p$-brane charge, by
performing the rescalings \bea &&m\longrightarrow \ep^{\tilde d} m\
,\qquad \sinh\a \longrightarrow \ep^{-{{\tilde d}\over 2}}\, \sinh\a\
,\nn\\ &&r\longrightarrow \ep\, r\ , \qquad x^\mu\longrightarrow
\ep^{1-\td d/2}\, x^\mu\ , \qquad \ell_i \rightarrow \ep\, \ell_i\ ,
\eea and then sending $\ep$ to zero.  We find that the metric becomes
$ds^2 =  \ep^{a^2/2}\, d\td s^2$, where $a$ is given under
(\ref{dsinglelag}) and the metric $d\td s^2$ is given by
\bea
d\td s^2_{D} &=& \wtd \Delta^{\ft{\td d}{D-2}}\, e^{\td d\varphi}\,
\Big[-(H_1\cdots H_N)^{-\ft{d-2}{d-1}}\, f\, dt^2\nn\\
&& + (H_1\cdots H_N)^{\ft{1}{d-1}}\Big(
(\fft{\td d}{d}\,g\rho)^{\ft{(D-2)a^2}{2\td d}}\, f^{-1}\,  d\rho^2 +
\rho^2\, d\vec y\cdot d\vec y\Big)\Big]\nn\\
&&+g^{-2}\, \wtd \Delta^{-\ft{d}{D-2}}\, e^{-d\varphi}\,
\sum_{i=1}^N X_i^{-1}\, \Big(d\mu_i^2 + \mu_i^2 (d\phi_i +
g\, A^i)^2\Big)\ ,
\eea
where
\bea
&&g\,\rho =(d/\td d)\, (g\, r)^{\td d/d}\ ,\qquad \vec y= g\, (\td
d/d)\, \vec x\ ,\nn\\
&& g^{-\td d}=2m\, \sinh^2\a\ ,\qquad \mu=2m\, (d/\td d)^{d-2}
\, g^{2+\td d-d}\ ,
\eea
and
\bea
&&f=-\fft{\mu}{\rho^{d-2}} + (\td d/d)^2 g^2\, \rho^2\, (H_1\cdots
H_N)\ ,\qquad
X_i=(H_1\cdots H_N)^{\ft{d}{(d-1)\td d}}\, H_i^{-1}\ ,\nn\\
&&\wtd \Delta = \fft{\sum_i X_i\,\mu_i^2}{(X_1\cdots X_N)^2}\ ,\qquad
e^{-\ft{2\td d}{a^2}\varphi} = (\td d/d)\, g\, \rho\ .\qquad
A^i=\fft{1-H_i^{-1}}{g\, \ell_i\, \sinh\a}\, dt\ .
\eea
It follows that the $(d+1)$ dimensional metric becomes
\be
ds_{d+1}^2 =-(H_1\cdots H_N)^{-\ft{d-2}{d-1}}\, f\, dt^2 +
(H_1\cdots H_N)^{\ft{1}{d-1}}\Big(
e^{-(D-2)\, \varphi}\, f^{-1} \, d\rho^2 +
\rho^2\, d\vec y\cdot d\vec y\Big)\ .
\ee
The Einstein-frame metric is given by $ds_E^2 = e^{-\fft{(D-2)}{(d-1)}
\varphi}\, ds_{d+1}^2$. 
This is the metric of an $N$-charge black hole in a domain-wall background.
In the case when $a=0$, the domain wall specialises to AdS$_{d+1}$.

\section{Conclusions}

    In this paper, we have constructed the non-linear ans\"atze for
the spherical dimensional reduction of type IIB supergravity on $S^5$,
and eleven-dimensional supergravity on $S^7$ and $S^4$, in the case
where we restrict to the abelian subgroups of the full $SO(6)$, $SO(8)$
and $SO(5)$ gauge groups.  In this way, we have shown how the 
gauged theories in $D=5$, $D=4$ and $D=7$ that are relevant for
constructing charged AdS black holes can be embedded into type IIB
supergravity or eleven-dimensional supergravity.

   As a matter of fact, in order to work out what the non-linear
metric and field-strength ans\"atze should be, we got many hints by
looking at the detailed forms of the lower-dimensional configurations
that one obtains by making the appropriate sphere compactifications of
the near-horizon limits of the corresponding spinning D3-brane,
M2-brane or M5-brane, and comparing the results with the AdS
black-hole solutions in $D=5$, $D=4$ and $D=7$.  The key step in being
able to extract general results for the reduction ans\"atze from these
specific solutions is that one must first establish that the various
components of the higher-dimensional metrics and field strengths can
in fact be expressed in a generic way in terms of the fields of the
lower-dimensional theory, together with the coordinates of the
compactifying sphere.  Only by doing this can one then ``kick away the
ladder,'' and extract general results, independent of any {\it
specific} solution, for how an {\it arbitrary} solution of the
lower-dimensional equations of motion can be embedded in the
higher-dimensional theory.  Luckily, the multi-charge AdS black-hole
solutions are general enough that they provided many clues that were
helpful in deducing what the full ans\"atze should be.

    Rather general, although highly implicit, results had previously
been given for the metric ansatz for the full $SO(8)$ reduction of
$D=11$ supergravity on $S^7$ \cite{deWitnicolaiwarner,deWitnicolai}.
In principle, it should be possible to verify that the $U(1)^4$
truncation of the general case is in agreement with our result
(\ref{s7metred}) for the $U(1)^4$ gauged theory.  In practice, however,
the implicit nature of the general expressions in
\cite{deWitnicolaiwarner,deWitnicolai} makes the comparison rather
difficult. The situation regarding the 4-form field strength is even
less clear, and only results of an even more implicit nature have been
previously presented.  Even less has been given previously for the
other examples, namely the $S^5$ and $S^4$ reductions.  It should be
emphasised that it is the handling of the scalar fields in the various
ans\"atze that presents the major challenges in performing spherical
Kaluza-Klein reductions.

   As far as we are aware, therefore, the results in this paper
provide the first explicit examples of compactifications of $D=11$
supergravity on $S^4$ and $S^7$, and type IIB supergravity on $S^5$,
in which sets of lower-dimensional massless fields that include scalars are
embedded in the reduction ans\"atze.  Importantly, we showed that the
equivalence between the corresponding gauged supergravities and the
compactified higher-dimensional theory is at the level of the
equations of motion, rather than at the level of the effective
Lagrangian.  Furthermore, the fact that one is able at all to read off
sensible lower-dimensional equations of motion depends crucially on
conspiracies between the contributions of the ans\"atze for the
higher-dimensional metric and antisymmetric tensor.  (Without such
conspiracies, one would not get a clean factorisation of lower-dimensional
equations of motion multiplied by overall sphere-dependent factors.)
This emphasises that non-trivial spherical reductions, in which
Kaluza-Klein gauge fields and scalars are retained, make sense only in
the context of certain very special higher-dimensional theories.  All
the known examples of such theories are supergravities. 

  Having found the non-linear Kaluza-Klein ans\"atze, we were able to
provide an explicit demonstration of how the multi-charge AdS black
holes in gauged $D=5$, $D=4$ and $D=7$ supergravities that have
toroidal horizons are embedded
in type IIB supergravity or $D=11$ supergravity, and to show that the
higher-dimensional solutions are precisely the near-horizon decoupling
limits of spinning D3-branes, M2-branes and M5-branes.  (Previous
partial results for a single-charge AdS$_5$ black hole appeared in
\cite{Cveticgubser1}, and results for the special case of
Reissner-Nordstr\"om AdS$_5$ and AdS$_4$ black holes appeared in
\cite{Chamblin}.)

  The results that we have obtained have also opened the door to the
study of the embedding into M-theory or string theory of other
solutions of gauged supergravities in $D=4$, $D=5$ and $D=7$ in
dimensions; for example AdS black holes with other topologies (such as
spheres or hyperbolic spaces), strings, domain walls, {\it etc}.
These solutions could in turn provide novel information about other
possible distortions of the spherical compactifications (not only
those related to rotations), and thus provide new insights into
strongly-coupled gauged theories via the AdS/CFT correspondence.

\section*{Acknowledgements}

We made extensive use of the TTC tensor-manipulation package.
J.T.L.~and C.N.P.~thank the University of Pennsylvania for
hospitality, and J.T.L. thanks Texas A\&M University for hospitality.  

\section*{Appendices}

\appendix

\section{Single-charge rotating $p$-branes}

    In this appendix, we present, for convenience, some general
results for rotating $p$-branes in arbitrary dimensions, supported by
a single $(p+2)$-form charge.  This are all straightforwardly obtained
by diagonally oxidising the rotating black holes constructed in
\cite{Cveticyoum3}.

   Single-charge $p$-branes in supergravity theories are solutions of
the Lagrangian
\be e^{-1}{\cal L} = R - \ft12 (\del\phi)^2 - \ft1{2n!}\, e^{a\phi}\,
(F_\n)^2 \ ,\label{dsinglelag}
\ee
where $F_\n=dA_{\sst{(n-1)}}$ and $a^2 = 4 -2 (n-1)(D-n-1)/(D-2)$
\cite{dufflublack}. In
this appendix, we obtain rotating $p$-brane solutions.  The Lagrangian
(\ref{dsinglelag}) admits an electric $(d-1)$-brane with $d=n-1$ or a
magnetic $(d-1)$-brane with $d=D-n-1$.  We shall consider only the
electric solution here, since the magnetic one can be viewed as an
electric solution of the dual $(D-n)$-form field strength
$F_{\sst{(D-n)}}$.  The rotating $p$-brane can be dimensionally
reduced on its world-volume spatial coordinates, to give rise to
single-charge rotating black holes, which were obtained in
\cite{Cveticyoum3}.  Conversely, it is a straightforward procedure
dimensionally to oxidise the rotating black hole solutions in
\cite{Cveticyoum3} to give the rotating $p$-branes in higher
dimensions.  We shall use this approach to obtain general
single-charge rotating $p$-branes in this appendix.

         Introducing a dual parameter $\td d=D-d-2=D-n-1$, the dimension
of the transverse space is $\td d+2$.  It follows that the foliating
spheres of the transverse space have dimension $\td d +1$.  There are
two cases arising, depending on whether $\td d$ is even or odd.

\bigskip\bigskip
\noindent{\underline{{\it Case 1:}\ \ $\td d +2 = 2N$}}
\bigskip

       In this case, there are $N$ angular momenta $\ell_i$, with
$i=1,2,\ldots, N$.  We find that the metric of the rotating
$(n-2)$-brane solution to the equations following from
(\ref{dsinglelag}) is
\bea
ds_{D}^2&=& H^{-\ft{\td d}{D-2}}\Big(-(1 - \fft{2m}{r^{\td d}\Delta})\,
dt^2 + d\vec x\cdot d\vec x\Big) +
H^{\ft{d}{D-2}}\Big[\fft{\Delta\, dr^2}{H_1\cdots H_N -2m\,
r^{-\td d} }\nn\\
&&+r^2 \sum_{i=1}^N H_i(d\mu_i^2 + \mu_i^2\, d\phi_i^2) -
\fft{4m\cosh\a}{r^{\td d}\, H\, \Delta}\, dt\,
(\sum_{i=1}^N \ell_i\, \mu_i^2\, d\phi_i)\nn\\
&&+\fft{2m}{r^{\td d}\, H\, \Delta}\,
(\sum_{i=1}^N \ell_i\, \mu_i^2\, d\phi_i)^2 \Big]\ ,
\label{tddeven}
\eea
where the functions $\Delta$, $H$ and $H_i$ are given by
\bea
&&\Delta = H_1\cdots H_N\, \sum_{i=1}^N \fft{\mu_i^2}{H_i}
\ ,\qquad H= 1 + \fft{2m\, \sinh^2\a}{r^{\td d}\, \Delta}\ ,
\nn\\
&&H_i = 1 + \fft{\ell_i^2}{r^2}\ ,\qquad i=1,2,\ldots, N\ .
\eea
The dilaton $\phi$ and gauge potential $A_{\sst{(n-1)}}$ are given by
\be
e^{2\phi/a} = H\ ,\qquad
A_{\sst{(n-1)}} = \fft{1-H^{-1}}{\sinh\a}\Big(\cosh\a \, dt +
\sum_{i=1}^N \ell_i\, \mu_i^2\, d\phi_i\Big )\wedge d^{n-2}x\ .
\ee
The $N$ quantities $\mu_i$, as usual, are subject to the constraint
$\sum_i \mu_i^2=1$.  One can parameterise the $\mu_i$ in terms of
$(N-1)$ unconstrained angles.  A common choice is
\bea
\mu_i &=& \sin\psi_i\, \prod_{j=1}^{i-1}\cos\psi_j\ ,
\qquad i\le N-1\ ,\nn\\
\mu_N &=& \prod_{j=1}^{N-1}\cos\psi_j\ .
\eea
Note that $\prod_{j=1}^n \cos\psi_j$ is defined to be equal to 1 if
$n\le 0$.

\bigskip\bigskip
\noindent{\underline{{\it Case 2:}\ \ $\td d +2 = 2N+1$}}
\bigskip

   Here, the solution has the same form as in Case 1, but with the
range of the index $i$ extended to include 0.  (Note that our variable
$\mu_0$ is called $\a$ in \cite{mype}.)  However, there is no angular
momentum parameter or azimuthal coordinate associated with the extra
index value, and so $\ell_0=0$ and $\phi_0=0$.  Otherwise, all the
formulae in Case 1 are generalised simply by extending the summation
to span the range $0\le i\le N$.  Of course $H_0=1$ as a consequence
of $\ell_0=0$.

\section{$D=4$, $N=2$ gauged supergravity}

The $SO(8)$ gauged $N=8$ supergravity in four dimensions was obtained
in \cite{deWit1,deWit2} by gauging an $SO(8)$ subgroup of the global
$E_7$ symmetry group of \cite{Cremmer1,Cremmer2}.  To avoid some of
the complications of non-abelian gauge fields, one may consider a
truncation of this model to $N=2$, for which the bosonic sector
comprises the metric, four commuting $U(1)$ gauge potentials, three
dilatons and three axions.  In the absence of axions, this truncation
was obtained in \cite{Duffliu} by working in the symmetric gauge for
the 56-bein and incorporating three real scalars.  As was noted, there
is a straightforward generalization of the scalar ansatz to allow for
complex scalars.  Taking into account the $E_7$ self-duality condition
$\overline\phi^{ijkl}=\phi_{ijkl}={1\over4!}\epsilon_{ijklmnpq}\phi^{ijkl}$,
the scalar ansatz of \cite{Duffliu} may be generalized as:
\begin{equation}
\overline\phi^{ijkl}=\phi_{ijkl}=\sqrt{2}[
\Phi^{(1)}\epsilon^{(12)}+\Phi^{(2)}\epsilon^{(13)}+\Phi^{(3)}
\epsilon^{(14)}+
\overline\Phi^{(1)}\epsilon^{(34)}+\overline\Phi^{(2)}\epsilon^{(24)}+
\overline\Phi^{(3)}\epsilon^{(23)}]_{ijkl},
\end{equation}
where we follow the notation and conventions of \cite{Duffliu} (including the
definition of $SO(8)$ index pairs).  Note that the three complex scalars
may be parameterised in terms of their magnitudes and phases as
$\Phi^{(i)}=\phi^{(i)}e^{i\theta^{(i)}}$.

    Here, we shall consider the full $N=2$ truncation, where the three
axions are included as well as the other fields. In fact the structure
of the potential is little changed.  We find that the Lagrangian
including the axions may be written in the form
\begin{equation}
e^{-1}{\cal L}_4= R - \half\sum_i\left((\partial\phi^{(i)})^2
+\sinh^2\phi^{(i)}(\partial\theta^{(i)})^2\right)
- \half(F_{\mu\nu}^{(\alpha)+}{\cal M}_{\alpha\beta}
F^{(\beta)+\mu\nu}+{\rm h.c.})-g^2V\ ,\label{d44clag}
\end{equation}
where the potential is given simply by
\begin{equation}
V=-8(\cosh\phi^{(1)}+\cosh\phi^{(2)} +\cosh\phi^{(3)})\ .
\end{equation}
The complex symmetric scalar matrix ${\cal M}$ is quite complicated, and
incorporates all three complex scalars $\Phi^{(\alpha)}$ in a symmetric
manner; this is presented below.

In terms of the $N=2$ truncation, the three complex scalars each
parameterise an $SL(2;R)/SO(2)$ coset.  This may be made explicit by
performing the change of variables $(\phi^{(i)},\theta^{(i)})
\to(\varphi_i,\chi_i)$:
\begin{eqnarray}
\cosh\phi^{(i)}&=&\cosh\varphi_i+\half\chi_i^2\, e^{\varphi_i}\ , \nonumber\\
\cos\theta^{(i)}\sinh\phi^{(i)}&=&\sinh\varphi_i
-\half\chi_i^2\, e^{\varphi_i}\ ,\nonumber\\
\sin\theta^{(i)}\sinh\phi^{(i)}&=&\chi_i \, e^{\varphi_i}\ .
\end{eqnarray}
Defining the dilaton-axion combinations
\begin{equation}
A_i = 1 + \chi_i^2\,  e^{2\varphi_i}\ ,
\end{equation}
as well as
\begin{eqnarray}
B_1&=&\chi_2\, \chi_3 \, e^{\varphi_2+\varphi_3} +i\chi_1\,
e^{\varphi_1}\ ,\nonumber\\
B_2&=&\chi_1\, \chi_3 \, e^{\varphi_1+\varphi_3} +i\chi_2\,
e^{\varphi_2}\ ,\nonumber\\
B_3&=&\chi_1\, \chi_2 \, e^{\varphi_1+\varphi_2} +i\chi_3 \,
e^{\varphi_3}\ ,
\end{eqnarray}
we finally obtain the bosonic Lagrangian
\begin{equation}
e^{-1}{\cal L}_4=R - \half\sum_i\left((\partial\varphi_i)^2
+e^{2\varphi_i}(\partial\chi_i)^2\right)
- \half(F_{\mu\nu}^{(\alpha)+}{\cal M}_{\alpha\beta}
F^{(\beta)+\mu\nu}+{\rm h.c.})-g^2V\ .
\end{equation}
The potential $V$ is now given by
\begin{equation}
V=-8\sum_i\left(\cosh\varphi_i + \half\chi_i^2\, e^{\varphi_i}\right)
\ ,\label{eq:sl2pot}
\end{equation}
and the scalar matrix is
\begin{equation}
{\cal M}={1\over D}\left[\matrix{
e^{-\lambda_1} &e^{\varphi_1}B_1 &e^{\varphi_2}B_2 &e^{\varphi_3}B_3\cr
e^{\varphi_1}B_1 &e^{-\lambda_2}A_2A_3 &-e^{-\varphi_3}A_3 B_3
&-e^{-\varphi_2}A_2B_2\cr
e^{\varphi_2}B_2 &-e^{-\varphi_3}A_3B_3 &e^{-\lambda_3}A_1A_3
&-e^{-\varphi_1}A_1B_1\cr
e^{\varphi_3}B_3 &-e^{-\varphi_2}A_2B_2 &-e^{-\varphi_1}A_1B_1
&e^{-\lambda_4}A_1A_2}\right]\ ,
\end{equation}
where
\begin{equation}
D=1+\chi_1^2\, e^{2\varphi_1} +\chi_2^2\, e^{2\varphi_2} +\chi_3^2\,
e^{2\varphi_3}
-2i\, \chi_1\, \chi_2\, \chi_3\, e^{\varphi_1+\varphi_2+\varphi_3}\ .
\end{equation}
The scalar combinations $\{\lambda\}$ are defined as in \cite{Duffliu}:
\begin{eqnarray}
\lambda_1&=&           -\varphi_1-\varphi_2-\varphi_3\ ,\nonumber\\
\lambda_2&=&           -\varphi_1+\varphi_2+\varphi_3\ ,\nonumber\\
\lambda_3&=&\hphantom{-}\varphi_1-\varphi_2+\varphi_3\ ,\nonumber\\
\lambda_4&=&\hphantom{-}\varphi_1+\varphi_2-\varphi_3\ .
\label{eq:lambdas}
\end{eqnarray}

While this $N=2$ truncation of the $N=8$ theory essentially treats all
four $U(1)$ gauge fields equally, it was noted that one can make
contact with the theory obtained by reduction of a closed string on
$T^2$ through dualisation of two of the gauge fields.  To be specific,
we dualise $F_{\mu\nu}^{(2)}$ and $F_{\mu\nu}^{(4)}$, which singles
out the dilaton-axion pair $S=\chi_2 + i e^{-\varphi_2}$.  After an
additional field redefinition $S\to -1/S$, we obtain the bosonic
Lagrangian
\begin{eqnarray}
e^{-1}{\cal L}_4^{\rm dualized} &=&R -\half(\partial\varphi_2)^2
-\half e^{2\varphi_2}(\partial\chi_2)^2
+ \ft18\Tr (\partial ML \partial ML)
-g^2V\nonumber\\
&&\qquad
-\ft14 e^{-\varphi_2} F^T(LML)F - \ft14 \chi_2 \, F^TL{*F}\ ,
\label{eq:dlag}
\end{eqnarray}
where the potential is still given by (\ref{eq:sl2pot}).
The scalar matrix $M$ is given in terms of the
$SL(2;R)\times SL(2;R)$ vielbein
\begin{equation}
{\cal V}=e^{\varphi_3/2}\left[\matrix{1&-\chi_3\cr 0&e^{-\varphi_3}}\right]
\otimes e^{\varphi_1/2}\left[\matrix{1&-\chi_1\cr 0&e^{-\varphi_1}}\right],
\end{equation}
by $M={\cal V}^T{\cal V}$,
and the gauge fields have been arranged in the particular order
\begin{equation}
F_{\mu\nu}=[\matrix{F_{\mu\nu}^{(3)}&\widetilde F_{\mu\nu}^{(4)}&
\widetilde F_{\mu\nu}^{(2)}&-F_{\mu\nu}^{(1)}}]^T.
\end{equation}
Finally, $L=\sigma^2\otimes\sigma^2$ satisfies $L^2=I_4$ where
$\sigma^2$ is the standard Pauli matrix.  It is worth mentioning that
the pure scalar Lagrangian can be expressed as
\be
e^{-1} {\cal L}_{\rm scalar} = \sum_{i=1}^3 \Big[-\ft12 \tr \del {\cal M}_i
\del {\cal M}^{-1}_i + 4g^2 \tr {\cal M}_i \Big]\ ,
\ee
where ${\cal M}_i = {\cal V}_i^T\, {\cal V}_i$ and ${\cal V}_i$ is given by
\be
{\cal V}_i = e^{\varphi_i/2}\left[\matrix{1&-\chi_i\cr
0&e^{-\varphi_i}}\right]
\ .
\ee

We see that, save for the potential, the dualise Lagrangian is
indeed of the form obtained from $T^2$ compactification from six
dimensions.  In this case, two of the $SL(2;R)$'s now correspond to
$T$-dualities while the third corresponds to $S$-duality.  Note that
the initial choice of which two field strengths to dualise has
determined which of the three dilaton-axion pairs $(\varphi_i,\chi_i)$ is
to be identified with the strong-weak coupling $SL(2;R)$.

  Having shown that the bosonic Lagrangian is considerably simplified
by dualising to the field variables that arise in the $T^2$ reduction,
we may re-express the result (\ref{eq:dlag}) in the more explicit
notation of \cite{lpsol,cjlp1}.  Thus the bosonic sector of the gauged
$U(1)^4$ theory may be written as
\bea
e^{-1}{\cal L}_4 &=& R -\ft12 (\del\vec\varphi)^2 -\ft12 e^{-\vec
a\cdot\vec\varphi}\, (\del\chi)^2 - \ft12 e^{\vec
a_{12}\cdot\vec\varphi}\, (F_{\1 12})^2 - \ft12 e^{\vec
b_{12}\cdot\vec\varphi}\, (\cF^1_{\1 2})^2 -g^2 V\nn\\
&& -\ft14 \sum_{i=1}^2\Big( e^{\vec a_i\cdot\vec\varphi}\, (F_{\2i})^2 +
e^{\vec b_i\cdot\vec\varphi}\, (\cF_\2^i)^2\Big) - \ft12 \chi\,
\ep^{\mu\nu\rho\sigma}\, F_{\mu\nu\, i}\, \cF^i_{\rho\sigma}\ ,
\eea
where the field strengths are given by
\bea
F_{\2 1} &=& dA_{\1 1} + dA_{\0 12} {\cal A}_\1^2\ ,\qquad
\cF^1_{\2} = d\cA^1_{\2} - d\cA^1_{\0 2}\, \cA_\1^2 \ ,\nn\\
F_{\2 2} &=& dA_{\1 2} - \cA^1_{\0 2}\, dA_{\1 1} -
dA_{\0 12}\, A_\1^1\ ,\qquad
\cF_\2^2 = d\cA_\1^2 \ .
\eea
Here $\chi$, $A_{\0 12}$ and $\cA^1_{\0 2}$ are the axions $\chi_2$,
$\chi_1$ and $\chi_3$, and the potential is given by (\ref{eq:sl2pot}).

The inclusion of the potential term in the gauged supergravity theory
breaks all three $SL(2;R)$ symmetries to $O(2)$, acting as $\tau\to
(a\tau+b) /(c\tau+d)$ where
\begin{equation}
\pmatrix{a&b\cr c&d} =
\pmatrix{\cos\alpha/2&\sin\alpha/2\cr-\sin\alpha/2&\cos\alpha/2}.
\label{o2rot}
\end{equation}
In terms of the original $(\phi,\theta)$ scalar variables in
(\ref{d44clag}), this $O(2)$ subgroup corresponds to
$\theta\longrightarrow \theta+\alpha$.  The $O(2)$ symmetry is,
however, sufficient for generating dyonic solutions.  Nevertheless, we
note that the fermionic sector and in particular the supersymmetry
transformations are not invariant under this symmetry of the bosonic
sector.  One manifestation of this particular situation is the fact
that magnetic black holes of this theory are not supersymmetric, even
though they may be extremal.  Furthermore, in a related note, while it
is clear that the dualisation procedure performed above runs into
difficulty in the full $N=8$ theory with non-abelian $SO(8)$ gauging,
the fermionic sector does not admit such a straightforward
dualisation, even in the $N=2$ abelian truncation.  This is easily
seen by the fact that the gravitini are necessarily charged under the
gauge fields and hence couple to the bare gauge potentials themselves.

\section{Calculation of the Ricci tensor for the reduction ans\"atze}

    Many of the calculations involved in the spherical reduction
ans\"atze in this paper are quite involved, and some of them are more
conveniently performed by computer.  However, some of them are quite
tractable by hand calculation.  Here, we present some useful results
for some of the curvature calculations for the metric ansatz, which
can be presented in a relatively compact form if further
specialisations are made, as discussed below.

The general Kaluza-Klein ansatz for odd sphere $S^{2k-1}$ reductions
of the $D$-dimensional metric may be expressed in the form
\begin{equation}
ds_{D}^2 = \wtd\Delta^{a}\, ds_d^2 +
\wtd\Delta^{-b}\, \sum_{i=1}^k X_i^{-1}\, \Big( d\mu_i^2 + \mu_i^2\,
(d\phi_i + A^i_\1)^2 \Big)\ ,
\label{eq:bigmet}
\end{equation}
where we have set the radius of $S^{2k-1}$ to unity.  There are $k-1$
scalar degrees of freedom parameterised by the $k$ quantities $X_i$
satisfying the constraint $\prod_{i=1}^kX_i=1$.  This form of the line
element encompasses both the $S^5$ reduction of type IIB supergravity and
the $S^7$ reduction of eleven dimensional supergravity.  As defined
previously, $\wtd\Delta = \sum_{i=1}^k X_i\mu_i^2$ and $\sum_{i=1}^k
\mu_i^2=1$.

In the absence of the gauge fields, this metric has a block diagonal form,
with the blocks corresponding to the $d$-dimensional spacetime, the $k-1$
direction cosines $\mu_i$ and the $k$ azimuthal rotation angles $\phi_i$.
The main difficulty in computing the curvature of (\ref{eq:bigmet})
lies in the fact that the $\mu_i$'s are constrained.  Nevertheless, we may
perform an asymmetric choice of using the first $k-1$ of them as actual
coordinates, while expressing $\mu_k$ as the constrained quantity
$\mu_k = (1-\sum_{i=1}^{k-1}\mu_i^2)^{1/2}$.

Since numerous terms are involved in the computation, it is imperative to
clarify our notation.  We denote the lower-dimensional spacetime
indices by $\mu,\nu,\ldots=0,1,\ldots,d-1$, the direction cosine
indices by $\alpha,\beta,\gamma,\ldots=1,2,\ldots,k-1$ and the azimuthal
indices by $i,j,\ldots=1,2,\ldots,k$.  Note that for instance
implicit sums over $\alpha$ always run over $k-1$ values, while sums
over $i$ always run over the full $k$ values.

Thus (with vanishing gauge fields) the $D$-dimensional metric may be
expressed in the form
\begin{equation}
G_{MN} = {\rm diag}[\wtd\Delta^a g_{\mu\nu},\wtd\Delta^{-b}
\hat g_{ij},\wtd\Delta^{-b}\wtd g_{\alpha\beta}]\ ,
\end{equation}
where $\hat g_{ij} = X_i^{-1}\mu_i^2\delta_{ij}$ is diagonal and
\begin{eqnarray}
\wtd g_{\alpha\beta}&=& X_\alpha^{-1}\delta_{\alpha\beta} + X_k^{-1}
\hat\mu_\alpha\hat\mu_\beta\ ,\nonumber\\
\wtd g^{\alpha\beta}&=&X_\alpha\delta_{\alpha\beta}-\wtd\Delta^{-1}
X_\alpha X_\beta\mu_\alpha\mu_\beta\ ,
\end{eqnarray}
with $\hat\mu_\alpha\equiv\mu_\alpha/\mu_k$.  Note that ${\rm
det}\,\wtd g_{\alpha\beta}=\wtd\Delta/\mu_k^2$.  As the $\mu_\alpha$
themselves are coordinates, this allows the use of expressions such as
$\partial_\alpha\mu_k = - \hat\mu_\alpha$ and
$\partial_\alpha\hat\mu_\beta=\mu_k^{-1}(\delta_{\alpha\beta}+\hat\mu_\alpha
\hat\mu_\beta)$.  In addition, all $\alpha,\beta,\ldots$ indices are raised
and lowered with the metric $\wtd g_{\alpha\beta}$.

Using this specific form of the metric $\wtd g_{\alpha\beta}$ and the fact
that ${\rm det}\,\hat g_{ij}=\prod_{i=1}^k\mu_i^2$, we find ${\rm det}\,
G_{MN}={\rm det}\, g_{\mu\nu}\, 
\wtd\Delta^{\kappa+2a}\prod_{i=1}^{k-1}\mu_i^2$ where the product
provides the measure over the internal $S^{2k-1}$.  Here
$\kappa=a(d-2)-b(2k-1)+1$ so that $\sqrt{-G}R\sim \sqrt{-g}\, 
\wtd\Delta^{\kappa/2}$.
Hence one expects $\kappa=0$ in order to prevent any $\wtd\Delta$ dependence
from appearing in front of the lower-dimensional Einstein term.  Indeed we
see that $\kappa$ vanishes for both the $S^5$ and the $S^7$
reductions considered in the text.  

We have only computed selected components of the full $D$-dimensional
Ricci tensor which are of interest in the Kaluza-Klein reduction.  While we
have used an asymmetrical parameterisation of the direction cosines, the
final results are symmetrical in all $k$ of the $\mu_i$'s.  For the
lower-dimensional components of $R_{MN}$ we find
\begin{eqnarray}
R_{\mu\nu}&=& R^{(d)}_{\mu\nu} 
-\half\kappa\wtd\Delta^{-1}\partial_\mu\partial_\nu\wtd\Delta
+\ft14((a+2)\kappa-(a+b)b(2k-1)+a+2b)\wtd\Delta^{-2}\partial_\mu\wtd
\Delta\partial_\nu\wtd\Delta\nonumber\\
&&+\ft14 (\partial_\mu\wtd g^{\alpha\beta}\partial_\nu \wtd g_{\alpha\beta}
-X_i^{-2}\partial_\mu X_i\partial_\nu X_i)\\
&&+g_{\mu\nu}[-\ft{a}{2}\wtd\Delta^{-1}\nabla^2\wtd\Delta
-\ft{a}{4}(\kappa-2)\wtd\Delta^{-2}\partial^\rho\wtd\Delta\partial_\rho
\wtd\Delta]\nonumber\\
&&+g_{\mu\nu}\wtd\Delta^{a+b}[-\ft{a}{4}(\kappa+2a+2b-3)\wtd\Delta^{-2}
\partial^\alpha\wtd\Delta\partial_\alpha\wtd\Delta
-\ft{a}{2}(\wtd\Delta^{-1}\wtd\nabla^2\wtd\Delta+\wtd\Delta^{-1}
\mu_i^{-1}\partial^\alpha\mu_i\partial_\alpha\wtd\Delta)]\ ,
\kern-3pt
\nonumber
\end{eqnarray}
where $R^{(d)}_{\mu\nu}$ denotes the Ricci tensor of the
$d$-dimensional spacetime metric $g_{\mu\nu}$.  
We have also determined the internal
Ricci components $R_{ij}$ and $R_{\alpha\beta}$ necessary for computing the
$D$-dimensional scalar curvature.  For the former we find
\begin{eqnarray}
R_{ij}&=&\hat g_{ij}\wtd\Delta^{-a-b}[\ft14\kappa\wtd\Delta^{-1}X_i^{-1}
\partial^\rho
\wtd\Delta\partial_\rho X_i+\ft{b}{2}\wtd\Delta^{-1}\nabla^2
\wtd\Delta+\ft{b}{4}(\kappa-2)\wtd\Delta^{-2}\partial^\rho\wtd\Delta
\partial_\rho\wtd\Delta]\\
&&+\hat g_{ij}[-\half(\kappa+2a+2b-1)\wtd\Delta^{-1}\mu_i^{-1}\partial^\alpha
\mu_i\partial_\alpha\wtd\Delta-\mu_i^{-1}\wtd\nabla^2\mu_i+\wtd\Delta^{-1}
(X_i\sum X-X_i^2)\nonumber\\
&&\qquad+\ft{b}{4}(\kappa+2a+2b-3)\wtd\Delta^{-2}\partial^\alpha
\wtd\Delta\partial_\alpha\wtd\Delta+\ft{b}{2}\wtd\Delta^{-1}\wtd\nabla^2
\wtd\Delta +\ft{b}{2}\wtd\Delta^{-1}\mu_l^{-1}\partial^\alpha\mu_l
\partial_\alpha\wtd\Delta]\nonumber
\end{eqnarray}
(no sum on $i$), while for the latter we have
\begin{eqnarray}
R_{\alpha\beta}&=&
\wtd\Delta^{-a-b}[\half \wtd g^{\gamma\delta}\partial^\rho
\wtd g_{\alpha\gamma}
\partial_\rho \wtd g_{\beta\delta}-\ft14\kappa\wtd\Delta^{-1}\partial^\rho
\wtd g_{\alpha\beta}\partial_\rho\wtd\Delta-\half\nabla^2\wtd g_{\alpha\beta}]
\nonumber\\
&&+\wtd g_{\alpha\beta}\wtd\Delta^{-a-b}[\ft{b}{4}(\kappa-2)\wtd\Delta^{-2}
\partial^\rho\wtd\Delta\partial_\rho\wtd\Delta+\ft{b}{2}\wtd\Delta^{-1}
\nabla^2\wtd\Delta]\nonumber\\
&&+\wtd R_{\alpha\beta}-\half(\kappa+2a+2b-1)\wtd\Delta^{-1}
\wtd\nabla_\alpha\wtd\nabla_\beta\wtd\Delta
-\mu_i^{-1}\wtd\nabla_\alpha\wtd\nabla_\beta\mu_i\\
&&-\ft14((b-2)\kappa+a(a+b)d
+(b-2)(2a+2b-1))\wtd\Delta^{-2}\partial_\alpha\wtd\Delta\partial_\beta
\wtd\Delta \nonumber\\
&&+\wtd g_{\alpha\beta}[\ft{b}{2}\wtd\Delta^{-1}\wtd\nabla^2\wtd\Delta
+\ft{b}{4}(\kappa+2a+2b-3)\wtd\Delta^{-2}\partial^\gamma\wtd\Delta
\partial_\gamma\wtd\Delta+\ft{b}{2}\wtd\Delta^{-1}\mu_i^{-1}\partial^\gamma
\mu_i\partial_\gamma\wtd\Delta] \ .\nonumber
\end{eqnarray}
Note that $\wtd R_{\alpha\beta}$ as well as the covariant derivatives
$\wtd\nabla_\alpha$ are defined with respect to the $k-1$ dimensional metric
$d\wtd s^2 = \sum_{i=1}^kX_i^{-1}d\mu_i^2$.
While these expressions are rather unwieldy, they simplify considerably in
both the $S^5$ and the $S^7$ reductions, as many of the coefficients take
on simple values.

Finally, by taking the trace of the above, we find the expression for the
$D$-dimensional curvature scalar
\begin{eqnarray}
R&=& \wtd\Delta^{-a}[R^{(d)}
-(\kappa+a)\nabla^\rho(\wtd\Delta^{-1}\nabla_\rho\wtd\Delta)
+\ft14(\partial^\rho \wtd g_{\alpha\beta}\partial_\rho \wtd g^{\alpha\beta}
-X_i^{-2}\partial^\rho X_i\partial_\rho X_i)\nonumber\\
&&\qquad+\ft14(-(\kappa+a)(\kappa-1)+2b-(a+b)b(2k-1))
\wtd\Delta^{-2}\partial^\rho\wtd\Delta\partial_\rho\wtd\Delta] \nonumber\\
&&+\wtd\Delta^b[\wtd R
-(\kappa+2a+b-1)(\wtd\Delta^{-1}\wtd\nabla^2\wtd\Delta
+\wtd\Delta^{-1}\mu_i^{-1}\partial^\alpha\mu_i\partial_\alpha\wtd\Delta)
\nonumber\\
&&\qquad+\ft14(-(\kappa+2a+b-1)(\kappa+2a+2b-5)-a(a+b)d)
\wtd\Delta^{-2}\partial^\alpha\wtd\Delta \partial_\alpha\wtd\Delta
\nonumber\\
&&\qquad-2\mu_i^{-1}\wtd\nabla^2\mu_i+\wtd\Delta^{-1}(\sum X)^2
-\wtd\Delta^{-1}\sum X^2]\ .
\end{eqnarray}
To make contact with the Kaluza-Klein reductions, we note that
explicit computation of $\wtd R$ yields
\begin{equation}
\wtd R=\wtd\Delta^{-1}[2\wtd\Delta^{-1}\sum X^3\mu^2-2\wtd\Delta^{-1}\sum X
\sum X^2\mu^2+(\sum X)^2-\sum X^2]\ ,
\end{equation}
where we have followed a shorthand notation of removing indices so that,
{\it e.g.}~$\sum X^3\mu^2\equiv\sum_{i=1}^kX_i^3\mu_i^2$.  Note that for the
special case of $k=3$, corresponding to $S^5$, not all of the above
quantities are independent.  As a result we find that this expression
simplifies to yield $\wtd R_{(k=3)}=2\wtd\Delta^{-2} X_1X_2X_3
=2\wtd\Delta^{-2}$.  Additionally, we often find the following identities
useful:
\begin{eqnarray}
\partial^\alpha\wtd\Delta\partial_\alpha\wtd\Delta
&=&-4[\wtd\Delta^{-1}(\sum X^2\mu^2)^2-\sum X^3\mu^2]\ ,\nonumber\\
\wtd\nabla^2\wtd\Delta&=&2[\wtd\Delta^{-2}(\sum X^2\mu^2)^2-\wtd\Delta^{-1}
\sum X^3\mu^2-\wtd\Delta^{-1}\sum X\sum X^2\mu^2+\sum X^2]\ ,\nonumber\\
\mu_i^{-1}\partial^\alpha\mu_i\partial_\alpha\wtd\Delta&=&-2[\wtd\Delta^{-1}
\sum X\sum X^2\mu^2-\sum X^2]\ ,\nonumber\\
\mu_i^{-1}\wtd\nabla^2\mu_i&=&\wtd\Delta^{-1}[
\wtd\Delta^{-1}\sum X\sum X^2\mu^2-(\sum X)^2]\ .
\end{eqnarray}

The $S^5$ reduction of type IIB supergravity discussed in section~2
corresponds to the choice of $d=5$ and $a=b=\half$.  In this case we obtain
\begin{eqnarray}
\wtd\Delta^{1/2}R_{(k=3)}^{(D=10)}\!\!\!&=&\!\!R^{(5)}
-\half\nabla^\rho(\wtd\Delta^{-1}
\nabla_\rho\wtd\Delta) -\ft14(-\partial^\rho\wtd g_{\alpha\beta}
\partial_\rho\wtd g^{\alpha\beta}
+X_i^{-2}\partial^\rho X_i\partial_\rho X_i
+\wtd\Delta^{-2}\partial^\rho\wtd\Delta\partial_\rho\wtd\Delta)\nonumber\\
&&+2(\wtd\Delta^{-1}+3\sum X^{-1})\ ,
\end{eqnarray}
where we have used the simplified expression for $\wtd R_{(k=3)}$ given
above.
On the other hand, the $S^7$ reduction of eleven dimensional supergravity,
given by the line element (\ref{s7metred}), corresponds to the choice of
$d=4$ and $a={2\over3}$, $b={1\over3}$.  The eleven-dimensional curvature
scalar is
\begin{eqnarray}
\wtd\Delta^{2/3}R_{(k=4)}^{(D=11)}\!\!\!&=&\!\!R^{(4)}
-\ft23\nabla^\rho(\wtd\Delta^{-1}
\nabla_\rho\wtd\Delta) -\ft14(-\partial^\rho\wtd g_{\alpha\beta}
\partial_\rho\wtd g^{\alpha\beta}
+X_i^{-2}\partial^\rho X_i\partial_\rho X_i
+\wtd\Delta^{-2}\partial^\rho\wtd\Delta\partial_\rho\wtd\Delta)\nonumber\\
&&-\ft23\wtd\Delta^{-2}(\sum X^2\mu^2)^2+\ft83\wtd\Delta^{-1}\sum X^3\mu^2
-\ft43\wtd\Delta^{-1}\sum X\sum X^2\mu^2\nonumber\\
&&+4(\sum X)^2-\ft{14}{3}\sum X^2\ .
\end{eqnarray}
The last two lines involve undifferentiated scalars, and is used in
(\ref{ricci1}).  Curiously, the scalar kinetic terms in both cases have an
identical structure save for a total derivative, and take on a standard
Kaluza-Klein appearance (since $X_i^{-2}\partial^\rho X_i\partial_\rho
X_i = - \partial^\rho\hat g_{ij}\partial_\rho\hat g^{ij}$).  Finally note
that the implicitly defined term
$\partial^\rho\wtd g_{\alpha\beta}\partial_\rho\wtd g^{\alpha\beta}$
may be evaluated to give
\begin{equation}
-\partial^\rho\wtd g_{\alpha\beta}\partial_\rho\wtd g^{\alpha\beta}
= X_i^{-2}\partial^\rho X_i\partial_\rho X_i
+\wtd\Delta^{-2}\partial^\rho\wtd\Delta\partial_\rho\wtd\Delta
-2\wtd\Delta^{-1}X_i^{-1}\partial^\rho X_i\partial_\rho X_i\mu_i^2\ .
\end{equation}

\end{document}